\documentclass[11pt]{article}

\usepackage[final]{acl}

\usepackage{times}
\usepackage{latexsym}

\usepackage[T1]{fontenc}
\usepackage[utf8]{inputenc}

\usepackage{microtype}

\usepackage{inconsolata}

\usepackage{graphicx}

\newcommand{\algname}{\emph{TimelyRAG}}
\newcommand{\benchname}{\emph{TimelyQABench}}
\usepackage{amsmath}
\usepackage{booktabs}
\usepackage{multirow}
\usepackage{pifont}
\usepackage{xcolor}
\usepackage{amssymb}
\usepackage{enumitem}
\usepackage{subcaption}

\usepackage[dvipsnames,table]{xcolor}
\newcommand{\heatG}[1]{%
  \ifdim #1pt<20pt  \cellcolor{ForestGreen!10!white}\strut #1\%%
  \else\ifdim #1pt<40pt \cellcolor{ForestGreen!20!white}\strut #1\%%
  \else\ifdim #1pt<60pt \cellcolor{ForestGreen!30!white}\strut #1\%%
  \else\ifdim #1pt<80pt \cellcolor{ForestGreen!40!white}\strut #1\%%
  \else                 \cellcolor{ForestGreen!50!white}\strut #1\%%
  \fi\fi\fi\fi
}

\newcommand{\heatR}[1]{%
  \ifdim #1pt<2.70pt \cellcolor{BrickRed!10!white}\strut #1%
  \else\ifdim #1pt<3.10pt \cellcolor{BrickRed!20!white}\strut #1%
  \else\ifdim #1pt<3.60pt \cellcolor{BrickRed!30!white}\strut #1%
  \else\ifdim #1pt<4.20pt \cellcolor{BrickRed!40!white}\strut #1%
  \else                 \cellcolor{BrickRed!50!white}\strut #1%
  \fi\fi\fi\fi
}

\title{\algname{}: Semantic-Temporal Hybrid Retrieval for Time-Critical Question Answering in Overlapping-Evolving Documents}

\author{
 \textbf{Youngeun Nam\textsuperscript{1}},
 \textbf{Joeun Kim\textsuperscript{1}},
 \textbf{Hwanjun Song\textsuperscript{1}},
 \textbf{Susik Yoon\textsuperscript{2}},
 \textbf{Jae-Gil Lee\textsuperscript{1}\thanks{Corresponding Author.}},
 \textbf{Byung Suk Lee\textsuperscript{3}}
\\
 \textsuperscript{1}KAIST,
 \textsuperscript{2}Korea University,
 \textsuperscript{3}University of Vermont
\\
 \texttt{\{youngeun.nam, je.kim, songhwanjun, jaegil\}@kaist.ac.kr}
 \\
 \texttt{susik@korea.ac.kr}
  \\
 \texttt{bslee@uvm.edu}
}

\begin{document}
\maketitle
\begin{abstract}
Although large language models\,(LLMs) and retrieval-augmented generation\,(RAG) have advanced open-domain question answering\,(QA), they remain unreliable when documents evolve through amendments.
Existing time-sensitive retrieval methods address only the \emph{disjoint}-evolving environment, where each update is an independent snapshot.
However, laws, policies, and regulations often operate in \emph{overlapping}-evolving environments, where amendments override earlier clauses while preserving most content, creating strong semantic overlap across versions.
We propose \algname{}, a retriever-agnostic framework that incorporates temporal distance into ranking to align queries with version-appropriate documents.
We also introduce \benchname{}, the first benchmark for regulation-heavy domains with overlapping-evolving challenges. 
Experiments show consistent gains, up to +28.6\% in nDCG@10, highlighting the importance of temporal reasoning for reliable QA over evolving documents.
All resources are available at \url{https://github.com/kaist-dmlab/TimelyRAG}.
\end{abstract}

\section{Introduction}
\label{sec:introduction}

The growing capabilities of LLMs have expanded open-domain QA systems into powerful tools for diverse domains.\,\cite{kamalloo2023evaluating, chen2024flexiqa, kim2024qpaug}. However, in the real world, LLM-based QA systems remain limited in \textit{time-critical} scenarios, where correctness depends not only on topical relevance but also on the validity of documents at query time and explicit temporal references in queries or documents. Amendments in domains such as law, policy, and institutional regulations preserve most content but revise key clauses, producing strong semantic overlap across successive versions. The discrepancy between document validity and a query's temporal context creates ambiguity in identifying supporting evidence, and addressing this fundamental challenge is essential in domains where decisions rely on precise, up-to-date regulatory knowledge.

Fine-tuning LLMs to reflect evolving knowledge is costly and difficult to sustain at scale\,\cite{chung2024scaling, brown2020language}. RAG\,\cite{lewis2020retrieval} provides a practical alternative by accessing updated document collections, but existing retrievers rely on \textit{lexical} or \textit{semantic} similarity\,\cite{robertson2009probabilistic, karpukhin2020dense, izacard2023unsupervised}. Without temporal awareness, they struggle to distinguish overlapping versions that share keywords but differ in validity. Existing time-sensitive methods rely on \textit{rough} temporal cues, typically year-level annotations, which are insufficient when amendments change clause-level conditions.

% \begin{figure*}[t]
%     \centering
%     % --- Subfigure A ---
%     \begin{subfigure}[t]{0.45\linewidth}
%         \centering
%         \includegraphics[width=\linewidth]{figures/fig_concept_v1.pdf}
%         \caption{Comparison of \textit{disjoint}- and \textit{overlapping}-evolving environments. Disjoint datasets are independent editions across time with no overlap, whereas overlapping datasets evolve through successive amendments, keeping most content while updating requirements.}
%         \label{fig:concept}
%     \end{subfigure}
%     \hfill
%     % --- Subfigure B ---
%     \begin{subfigure}[t]{0.53\linewidth}
%         \centering
%         \includegraphics[width=\linewidth]{figures/fig_example_v7.pdf}
%         \caption{Example of \textit{overlapping-evolving} environment. Queries from 2018 to 2025 require different valid versions of the same regulation. Since keywords such as ``English'' and ``Exam'' appear in all versions, traditional retrievers return every document without time awareness, whereas \algname{} aligns queries with the correct version.}
%         \label{fig:example}
%     \end{subfigure}

%     \caption{(a) Comparison of disjoint- and overlapping-evolving environments. 
%     (b) Example query with multiple document versions. 
%     Previous time-sensitive retrievers handle only disjoint settings, while \algname{} covers both.}
%     \label{fig:motivation_figure}
% \end{figure*}

\begin{figure}[t]
    \centering
    \includegraphics[width=\linewidth]{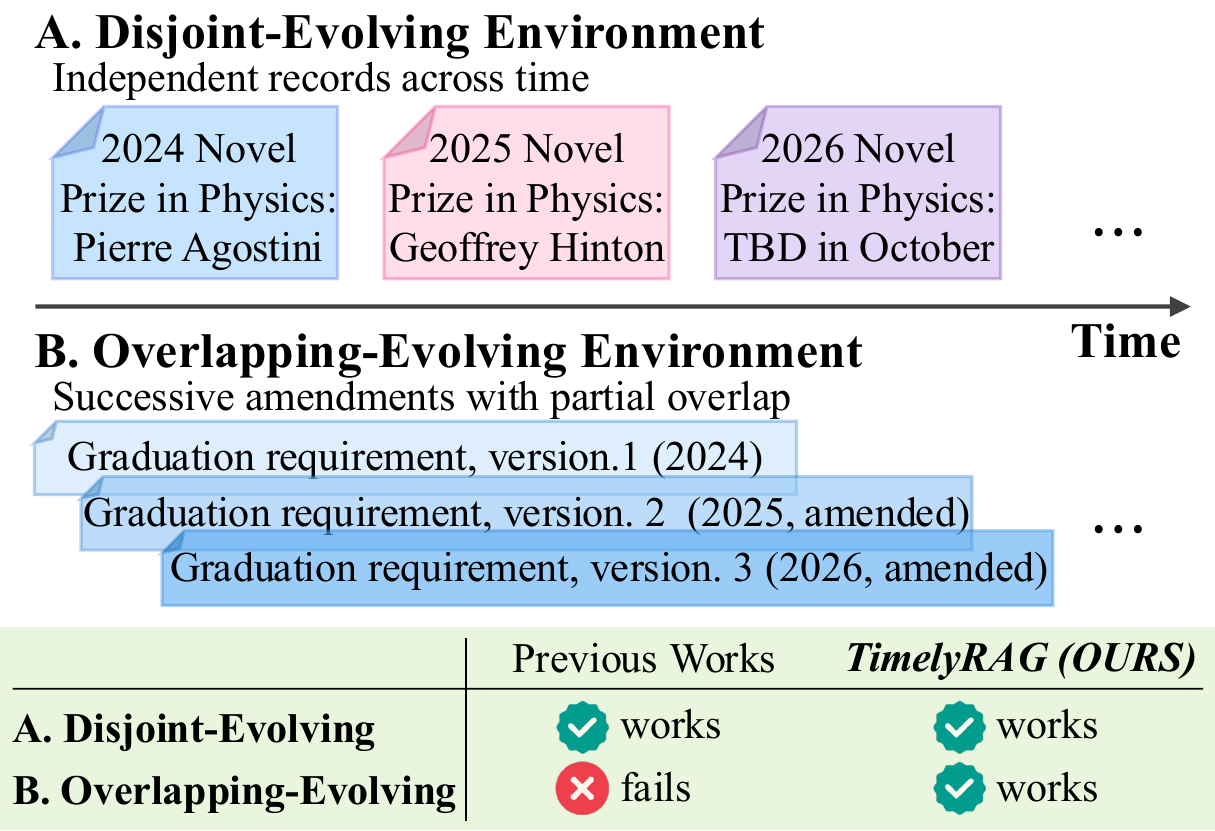}
    \caption{Comparison of \textit{disjoint}- and \textit{overlapping}-evolving environments. Disjoint datasets are independent editions, whereas overlapping ones retain most of the content with amendments. Prior retrievers mainly address disjoint settings; \algname{} supports both.}
    \label{fig:concept}
    \vspace{-0.3cm}
\end{figure}

We distinguish two temporal evolution environments. In the \textit{disjoint}-evolving environment, document editions are independent across time, as in yearly rankings or one-time event results. In the \textit{overlapping}-evolving environment, documents evolve through \emph{incremental amendments}, where most content remains but selected clauses change, as in graduation requirements or regulatory policies. Such corpora encode temporal validity through issuance dates, amendment dates, effective periods, or retroactive applicability clauses, either as metadata or formulaic text. Thus, retrieval must identify both the relevant document and the version whose clause is valid under the query's temporal context. Figure~\ref{fig:concept} contrasts the two environments.

\begin{figure}[t]
    \centering
    \includegraphics[width=\linewidth]{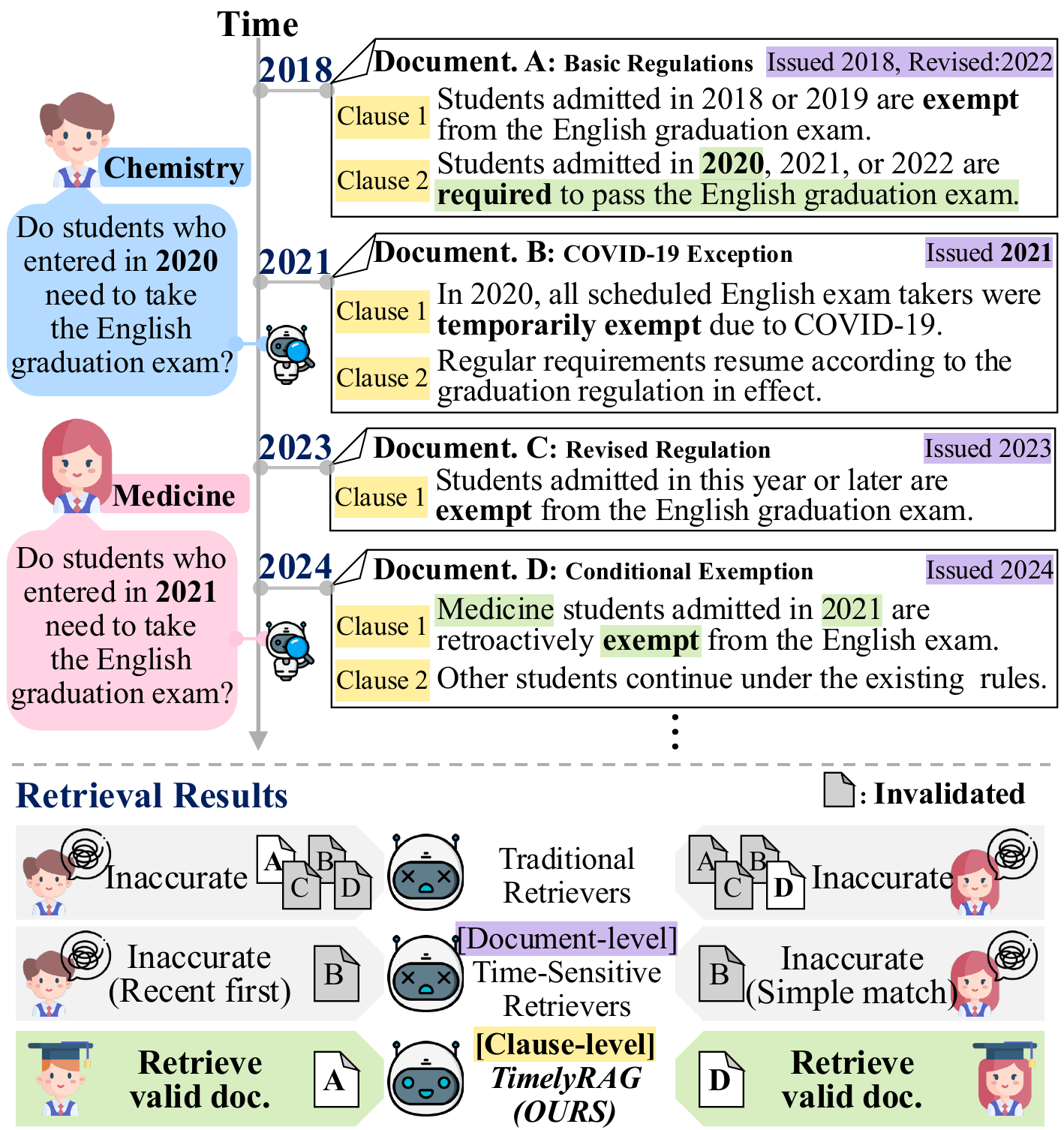}
    \caption{Example of \textit{overlapping-evolving} environment. Traditional retrievers return similar invalid versions, and document-level temporal retrievers can miss clause-level validity. \algname{} uses clause-level temporal cues to prioritize the version valid for the query.}
    \label{fig:example}
    \vspace{-0.3cm}
\end{figure}

Several benchmarks have been proposed for time-sensitive retrieval, including TimeQA\,\cite{chen2021dataset}, RealTime QA\,\cite{kasai2023realtime}, TS-Retriever\,\cite{wu2024time}, and CRAG\,\cite{yang2024crag}. These benchmarks evaluate temporal reasoning over encyclopedic facts, news freshness, timestamp-supervised retrieval, or dynamic domains. However, they mainly reflect disjoint-evolving settings, where timestamp matching or recency filtering is often sufficient. Overlapping-evolving environments involve amendments, where effective dates and clause-level conditions change. Correct retrieval thus requires matching query event times to the version's validity interval \emph{at the clause level}, rather than relying on a document-level date. Simple freshness heuristics fail when versions share keywords, insertion and effective dates diverge, or retroactive exceptions apply.

To address this gap, we introduce \algname{}, a retriever-agnostic framework that integrates temporal validity into ranking. \algname{} fuses semantic similarity with a \emph{temporal distance} that measures how the query time, or referenced event time, aligns with a candidate's effective interval.
The distance is minimal when the query time falls within a clause's effective interval and grows as it diverges. This clause-level modeling enables \algname{} to distinguish confusing amendments based on whether the required clause was valid. Figure~\ref{fig:example} contrasts \algname{} with document-level timestamp filtering. 
% Prior time-sensitive retrievers rely on simple timestamp matching, typically favoring the most recent or closest document date, even if the relevant clause is invalid. In contrast, \algname{} leverages the temporal distance to identify the version valid at the query time and to discard others where the clause is not in effect.

To evaluate \algname{}, we also construct \benchname{}, a benchmark tailored for \emph{overlapping-evolving environments} in regulation-focused domains. \benchname{} contains successive amendments with substantial semantic similarity across versions, requiring systems to retrieve temporally valid evidence, not merely similar documents. 

Overall, our key contributions consist of \algname{}, a retriever-agnostic semantic-temporal reranking framework; \benchname{}, the first benchmark centered on regulatory corpora with successive amendments; and experiments showing that \algname{} improves retrieval by up to +28.6\% in nDCG@10 and +19.1\% in Hit@10 across retrieval methods.
\section{Related Works}
\label{sec:related_work}

\subsection{Time-Aware RAG Methods}
Previous time-sensitive RAG methods incorporate temporal signals through timestamp filtering, recency-based scoring, or temporal supervision. TS-Retriever\,\cite{wu2024time} uses positive–negative contrastive learning with timestamp supervision, while MRAG\,\cite{zhang2024mrag} combines question processing, retrieval, summarization, and semantic-temporal hybrid ranking over temporally perturbed evidence. These methods highlight the importance of temporal information, but mainly target the \textit{disjoint-evolving environment}, where snapshots are independent and temporal reasoning reduces to selecting the correct edition. In contrast, the \textit{overlapping-evolving environment} requires distinguishing similar versions that differ in effective dates or scope, demanding temporal reasoning directly in the scoring process.

\subsection{Temporal KG Reasoning}
Temporal knowledge graph question answering (TKGQA) addresses time-dependent questions over structured facts. Representative work includes CronKGQA/CronQuestions\,\cite{saxena2021question}, EXAQT\,\cite{jia2021complex}, TempoQR\,\cite{mavromatis2022tempoqr}, and SubGTR\,\cite{chen2022temporal}, which model temporal constraints through temporal embeddings, compact subgraphs, or time-aware reasoning. These methods are complementary to \algname{} because they assume structured temporal facts in a knowledge graph, whereas \algname{} retrieves clause-valid versions from semantically overlapping textual amendments.

\subsection{Time-Sensitive QA Benchmarks}
Several benchmarks assess temporal reasoning in QA. TimeQA\,\cite{chen2021dataset} uses timestamped questions over Wikipedia, RealTime QA\,\cite{kasai2023realtime} evaluates continuously updated news questions, TS-Retriever\,\cite{wu2024time} adds timestamp supervision for retrieval, TimeR$^4$\,\cite{qian2024timer4} targets temporal knowledge graph QA, and CRAG\,\cite{yang2024crag} studies temporal consistency in financial and news domains. ChronoQA\,\cite{chen2025question} further studies time-sensitive RAG over large-scale news data with explicit and implicit temporal expressions. Despite these contributions, existing benchmarks rely on general-purpose corpora and provide limited coverage of incremental document evolution, primarily reflecting the \textit{disjoint-evolving environment}. % None of the existing benchmarks capture overlapping-evolving environments in which documents preserve most content across versions while specific clauses are amended, creating strong semantic similarity that complicates version-sensitive retrieval in policies, laws, and regulations.
\section{\algname{} Framework}
\label{sec:method}

\paragraph{Problem Formulation.} A \emph{time-critical query} is a query whose correct answer depends on the temporal validity of the retrieved evidence. In overlapping-evolving corpora, a candidate document can be topically relevant yet invalid because the clause needed to answer the query was not in effect under the query's temporal context.

Let $\mathcal{D}$ denote a collection of evolving documents and $d \in \mathcal{D}$ a candidate document. We consider four temporal signals for queries and documents:
\begin{itemize}[leftmargin=9pt, noitemsep]
    \item \textbf{Query Insert Time\,($Q_{IT}$)}: the time when the query was created or issued by the user.
    \item \textbf{Query Event Times\,($Q_{ET}$)}: the explicit temporal expressions mentioned within the query text\,(e.g., ``Which services became restricted after \textcolor{blue}{\texttt{January 2023}}?'')
    \item \textbf{Document Insert Time\,($D_{IT}$)}: the insertion timestamp of the document into the database.
    \item \textbf{Document Event Times\,($D_{ET}$)}: the temporal expressions extracted from the document content\,(e.g., ``Users who cancel a subscription within 14 days of activation, if purchased after \textcolor{blue}{\texttt{June 15, 2021}}, are eligible for a full refund.'').
\end{itemize}
These signals need not be equally available for all queries or documents. In particular, event times may be explicit, implicit, ambiguous, or missing. \algname{} therefore treats temporal information as a set of available signals and applies fallback-aware scoring when event-time evidence is incomplete.

% Formally, the answer to a time-sensitive query can be expressed as
% \begin{equation}
% \text{Answer}(Q) = f(Q, Q_{IT}, Q_{ET}, \mathcal{D}, D_{IT}, D_{ET}).
% \end{equation}
Let's suppose a retrieval function $f$, which 
% The retrieval function $f$ 
assigns each candidate document $d \in \mathcal{D}$ a relevance score that combines two components: the \emph{semantic similarity} between the query and the document, and the \emph{temporal compatibility} between their temporal attributes. For brevity, we denote the query-side temporal signals as $Q_T=(Q_{IT}, Q_{ET})$ and the document-side temporal signals as $D_T=(D_{IT}, D_{ET})$. The retrieval objective is to rank candidate documents by combining semantic relevance and temporal compatibility:
\begin{align}
f&(Q, Q_T, \mathcal{D}) \nonumber \\ 
&= \operatorname*{arg\,max}_{d \in \mathcal{D}} 
   \Big[ S(Q,d) + T(Q_T,D_T; d) \Big],
\end{align}
where $S(Q,d)$ denotes semantic similarity and $T(Q_T,D_T;d)$ denotes temporal compatibility.

\subsection{Semantic-Temporal Hybrid Retrieval}
\algname{} implements $f(Q,Q_T,\mathcal{D})$ as a two-stage retrieval pipeline. The first stage retrieves a candidate pool using a lexical or semantic retriever, while the second stage reranks the retrieved candidates using temporal compatibility. This design is retriever-agnostic: the first-stage retriever can be sparse, dense, or late-interaction based, and \algname{} only requires access to the resulting candidate set and temporal signals.

\begin{enumerate}[leftmargin=10pt]
    \item \textbf{Semantic retrieval stage}: A base retriever ranks all documents according to $S(Q,d)$ and retains the top-$N$ candidates.
    \item \textbf{Temporal re-ranking stage}: Candidate documents are re-ranked by combining $S(Q,d)$ with $T(Q_T,D_T;d)$, promoting documents whose relevant clauses are temporally valid under the query context.
\end{enumerate}

Because \algname{} is a reranking framework, it cannot recover a gold document that is absent from the first-stage candidate pool. Its goal is instead to resolve version-level ambiguity among semantically relevant candidates, which is the central difficulty in overlapping-evolving corpora.

\subsection{Stage 1: Semantic Retrieval}
\label{sec:stage1}
The first stage constructs a candidate pool using textual relevance only. Given a query $Q$, a base retriever ranks all documents by $S(Q,d)$ and retains the top-$N$ candidates, denoted by $\mathcal{C}_N(Q) \subset \mathcal{D}$. In our experiments, $N$ is substantially larger than the final evaluation cutoff, allowing temporal reranking over a sufficiently broad candidate pool. The final ranking after Stage 2 can be evaluated at different cutoffs, such as $k \in \{3,5,10\}$, depending on the downstream use case. For retrieval evaluation, we report standard metrics at $k=10$; for answer generation, the top-$k$ documents can be passed to the generator. We separately analyze candidate recall of $\mathcal{C}_N(Q)$ to distinguish first-stage coverage from temporal reranking effects.

\subsection{Stage 2: Temporal Re-Ranking}
\label{sec:stage2}
The second stage refines the candidate pool by incorporating temporal compatibility alongside semantic relevance.  
Document insert times are typically available from corpus metadata. In contrast, document event times are obtained from explicit metadata, rule-based temporal parsing, or offline temporal normalization of formulaic expressions in the document text. When event-time signals are missing or ambiguous, \algname{} falls back to available insertion-time signals and reduces reliance on uncertain temporal evidence through query-adaptive weighting.
For each candidate document $d \in \mathcal{C}_N(Q)$, the final score is defined as
\begin{align}
\label{eq:final_score}
\text{Score}(Q,d) &= (1-\alpha(Q))\,S(Q,d) \nonumber \\
                  &+ \alpha(Q)\,T(Q_T,D_T; d),
\end{align}
where $\alpha(Q) \in [0,1]$ balances semantic similarity and temporal compatibility.

\subsubsection{Temporal Signal Handling}

\algname{} distinguishes between insertion times and event times. Insertion times, such as $Q_{IT}$ and $D_{IT}$, indicate when a query or document version becomes available. Event times, such as $Q_{ET}$ and $D_{ET}$, indicate the time period to which the query or document clause refers. When event times are available, they are prioritized because they directly capture temporal validity. When an event time is missing, ambiguous, or low-confidence, \algname{} substitutes the corresponding insertion time only for the affected distance term rather than discarding the candidate.
We also enforce a causality constraint when insertion times are used; a document can support a query only if it was available before the query was issued, i.e., $D_{IT} \leq Q_{IT}$. This prevents future documents from being selected as evidence for past queries.

\subsubsection{Temporal Compatibility}
Temporal compatibility measures whether a candidate document is valid under the query's temporal context. For document event times, we represent clause validity as an interval $D_{ET}(d)=[s_d,e_d]$ when such interval information is available. Given a query event time $Q_{ET}$, the event-time distance is defined as zero if $Q_{ET}$ falls within the document's validity interval, and otherwise increases with the distance to the nearest boundary:
\begin{align}
&\Delta(Q_{ET},D_{ET}) \nonumber \\
& =
\begin{cases}
0, \quad\quad\quad\quad\quad\quad \text{if } Q_{ET} \in [s_d,e_d], \\
\min(|Q_{ET}-s_d|, |Q_{ET}-e_d|), \text{o/w.}
\end{cases}
\end{align}

When temporal information is represented as a point timestamp rather than an interval, this reduces to the absolute distance between the two timestamps. We then compute the overall temporal distance by combining available event-time and insertion-time alignments:
\begin{align}
\label{eq:temporal_distance}
\delta(Q,d) 
&= \Delta(Q_{ET},D_{ET})+\lambda_1 |Q_{IT}-D_{IT}| \nonumber \\
& +\lambda_2 \Delta(Q_{IT},D_{ET})+\lambda_3 |Q_{ET}-D_{IT}|,
\end{align}
where $\lambda_1,\lambda_2,\lambda_3$ control the contribution of auxiliary temporal alignments. If an event-time signal is missing, the corresponding term is replaced by the available insertion-time signal as described above.

The distance $\delta(Q,d)$ is normalized within the candidate pool $\mathcal{C}N(Q)$:
\begin{align}
& z(Q,d) \nonumber \\
&=
\frac{
\delta(Q,d) -
\min\limits_{d' \in \mathcal{C}_N(Q)} \delta(Q,d')
}{
\max\limits_{d' \in \mathcal{C}_N(Q)} \delta(Q,d')
\min\limits_{d' \in \mathcal{C}_N(Q)} \delta(Q,d')
+ \epsilon
},
\end{align}
where $\epsilon$ prevents division by zero. Finally, temporal compatibility is computed as
\begin{equation}
T(Q_T,D_T;d)=\exp(-\gamma z(Q,d)), \quad \gamma>0
\end{equation}
where $\gamma$ controls the temporal penalty strength.

\begin{figure*}[t]
    \centering
    \includegraphics[width=\linewidth]{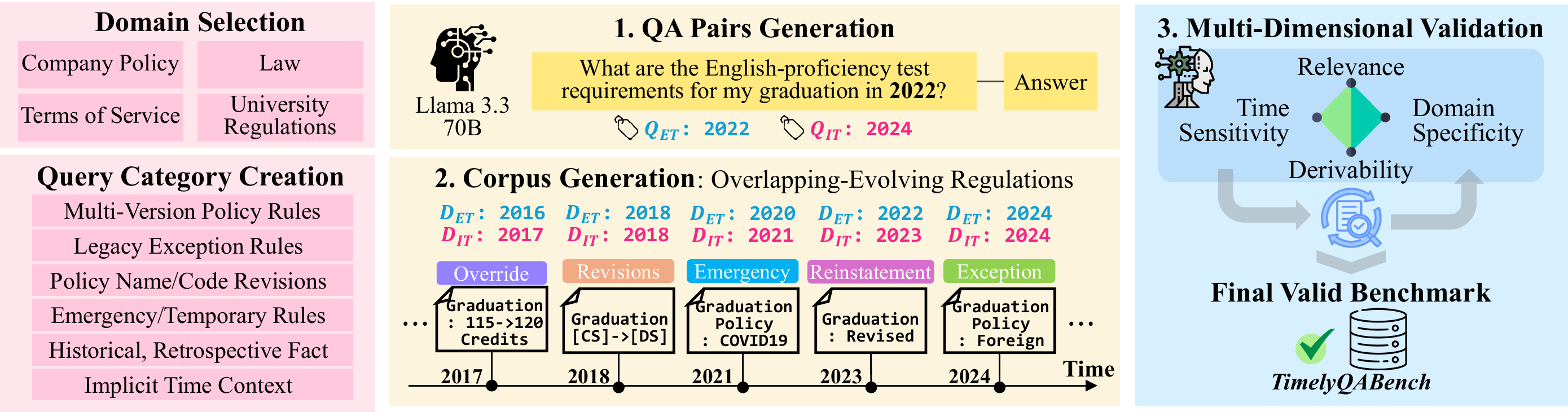}
    \caption{Overview of \benchname{} across four regulation-focused domains. Queries and documents are annotated with insertion timestamps, and validation uses four metrics to ensure reliable supervision.}
    \label{fig:validation}
    \vspace{-0.3cm}
\end{figure*}

\subsubsection{Balancing Semantic/Temporal Signals}
\label{sec:alpha_Q}
The coefficient $\alpha(Q)$ controls how strongly temporal compatibility affects the final ranking. Rather than assigning a fixed temporal weight to all queries, \algname{} adjusts $\alpha(Q)$ according to the temporal specificity of the query. Queries with explicit event times or fine-grained temporal expressions require stronger temporal discrimination, whereas queries with weak or absent temporal cues should rely more on semantic relevance.

We compute $\alpha(Q)$ as
\begin{equation}
\label{eq:final_alpha}
\alpha(Q) = \sigma\!\left(\gamma_1 \cdot \mathbf{1}_{\text{has ET}(Q)} 
+ \gamma_2 \cdot \text{Gran}(Q)\right),
\end{equation}
where $\mathbf{1}_{\text{has ET}(Q)}$ indicates whether the query contains an explicit event time,
$\text{Gran}(Q)$ encodes temporal granularity, and $\gamma_1,\gamma_2 \geq 0$ are weighting coefficients. This query-adaptive weighting prevents temporal signals from dominating when temporal evidence is weak, while emphasizing temporal validity when the query explicitly requires it.

% \paragraph{Scope of reranking.}
% \algname{} improves temporal validity within the candidate pool produced by the first-stage retriever. It does not replace candidate generation, nor does it recover documents that are never retrieved in Stage 1. This distinction is important because the effectiveness of any reranking method depends on the coverage of its first-stage candidate pool.

\begin{table*}[t]
    \centering
    \caption{Time-sensitive query categories, definitions, examples, and context in \benchname{}.}
    \resizebox{\linewidth}{!}{%
        \begin{tabular}{@{}llll@{}}
        \toprule
        \multicolumn{1}{c}{\textbf{Category}} & 
        \multicolumn{1}{c}{\textbf{Definition}} &
        \multicolumn{1}{c}{\textbf{Context}} &
        \multicolumn{1}{c}{\textbf{Examples}} \\
        \midrule
        Multi-Version Policy Rules &
          \begin{tabular}[c]{@{}l@{}}Policies that exist in multiple versions\\ with meaningful changes over time.\end{tabular} &
          \begin{tabular}[c]{@{}l@{}}Academic probation GPA requirement\\ changed from 2.0 to 2.5 after 2022.\end{tabular} &
          \begin{tabular}[c]{@{}l@{}}What GPA is required\\ to avoid academic probation?\end{tabular} \\ \midrule
        Legacy Exception Rules &
          \begin{tabular}[c]{@{}l@{}}Rules allowing certain entities to follow\\ older rules despite newer versions.\end{tabular} &
          \begin{tabular}[c]{@{}l@{}}Students enrolled before 2021 \\ follow the old graduation rules.\end{tabular} &
          \begin{tabular}[c]{@{}l@{}}Do students enrolled \\ in 2020 need a thesis?\end{tabular} \\ \midrule
        Policy Name/Code Revisions &
          \begin{tabular}[c]{@{}l@{}}Policies where course or program names\\ or codes are officially updated.\end{tabular} &
          \begin{tabular}[c]{@{}l@{}}“Data Security 301” was renamed \\ “Data Privacy 401” in 2024.\end{tabular} &
          \begin{tabular}[c]{@{}l@{}}Which new code covers\\ data privacy in 2024?\end{tabular} \\ \midrule
        Emergency/Temporary Rules &
          \begin{tabular}[c]{@{}l@{}}Temporary rules are issued in response\\ to emergencies or external events.\end{tabular} &
          \begin{tabular}[c]{@{}l@{}}Pass/fail grading was temporarily\\ allowed during COVID-19.\end{tabular} &
          \begin{tabular}[c]{@{}l@{}}Was pass/fail grading \\ allowed in Fall 2020?\end{tabular} \\ \midrule
        Historical, Retrospective Fact &
          \begin{tabular}[c]{@{}l@{}}Questions referring to how policies\\ existed at a specific time in the past.\end{tabular} &
          \begin{tabular}[c]{@{}l@{}}Tuition fees increased from \$5,000 \\ in 2015 to \$6,000 in 2018.\end{tabular} &
          \begin{tabular}[c]{@{}l@{}}What was the tuition fee \\ for students in 2015?\end{tabular} \\ \midrule
        Implicit Time Context &
          \begin{tabular}[c]{@{}l@{}}Queries where the intended time must\\ be inferred from contextual cues.\end{tabular} &
          \begin{tabular}[c]{@{}l@{}}Encryption rules were revised\\ in 2023, replacing older policies.\end{tabular} &
          \begin{tabular}[c]{@{}l@{}}What are the encryption \\ rules for banks?\end{tabular} \\ \bottomrule
        \end{tabular}
    }
    \label{tab:query_category}
    \vspace*{-0.1cm}
\end{table*}

\section{\benchname{} Construction}
\label{sec:dataset}

Most time-sensitive QA benchmarks capture disjoint-evolving environments with independent snapshots. In contrast, regulation-heavy domains often evolve through overlapping amendments, where content largely persists but selected clauses change. We construct \benchname{} as a controlled, synthetic benchmark with strong semantic overlap, explicit amendment histories, and gold labels for the version valid under each query's temporal context.
As shown in Figure~\ref{fig:validation}, \benchname{} has three stages: regulatory QA generation, evolving-family corpus construction with hard negatives, and multi-dimensional validation.

\subsection{Domain-Specific QA Creation}
The first phase defines the temporal reasoning requirements to be tested. We focus on four regulation-intensive domains---legal regulations, university regulations, company policy, and terms of service---because they are authoritative sources that undergo frequent amendments and require strict temporal validity. These domains are challenging for retrieval, as minor wording or scope changes can alter the correct answer.

% Within these domains, we analyze real-world regulatory documents and identify six recurring reasoning patterns: exemptions, amendments, validity periods, event-based conditions, version overrides, and temporal scope restrictions. From this analysis, we construct a taxonomy of categories summarized in Table~\ref{tab:query_category}.
Within these domains, we define six categories in Table~\ref{tab:query_category}, covering common temporal reasoning requirements such as exemptions, amendments, validity periods, event-based conditions, version overrides, and scope restrictions. For example, legacy exception rules require retrieving an older clause that remains valid for a specific population, while emergency rules require identifying a clause valid only during a limited period.

Using Llama-3.3-70B\,\cite{grattafiori2024llama}, we generate Korean time-critical QA instances following these category definitions. Each instance contains a query, a query insertion time $Q_{IT}$, an answer, and gold document identifiers. The gold documents specify the minimal evidence required to derive the answer under the query's temporal context. This explicit linkage allows retrieval systems to be evaluated on topical relevance and on whether they retrieve the temporally valid version. We generate 3,000 QA instances for each domain, resulting in 12,000 instances in total.
Detailed generation prompts are provided in Appendix~\ref{sec:prompts}.

\subsection{Corpus Generation}
The second phase constructs the evidence corpus. For each policy theme, we create an evolving document family with multiple temporally ordered versions. Versions within the same family share the same policy topic and retain most content, while selected clauses are revised in eligibility, scope, effective period, exception conditions, or required actions. This design creates hard negatives, where non-gold documents are semantically similar to the gold document but temporally invalid for the query.

Each document version is assigned a document insertion time $D_{IT}$, and clauses include event-time expressions or validity intervals corresponding to $D_{ET}$. For example, an earlier version allowed ``refund requests within {\color{blue}30} days of purchase,'' while a later amendment shortens the period to ``{\color{blue}14} days'' or restricts eligibility to annual subscriptions. These changes mirror amendment processes where new provisions override or qualify earlier ones while preserving surrounding text.

% we generate $10$ temporally varying document versions per query, with each version assigned a distinct revision timestamp\,(\texttt{doc\_insert\_time}). These variants differ in precise clauses---such as eligibility periods, required conditions, or applicable scope---while preserving the majority of content. For example, an early version may state that ``refund requests must be filed within {\color{blue}30} days of purchase,'' whereas a later amendment may change this clause to ``within {\color{blue}14} days'' or restrict eligibility to ``annual subscriptions only.'' This design mirrors real amendment processes, where new provisions override prior ones.

% and it ensures that each query is paired with authoritative supporting passages (\texttt{gold\_docs}) as well as distractor passages\,(\texttt{non\_gold\_docs}) within the same evolving corpus.

Each query is associated with \texttt{gold\_docs}, the minimal document versions required to derive the correct answer. Other versions in the same evolving family are treated as temporally challenging \texttt{non\_gold\_docs}. We quantify semantic overlap across versions in Section~\ref{sec:benchmark_analysis}. Detailed corpus generation prompts are included in Appendix~\ref{sec:prompts}.

% the minimal set of documents required to derive the correct answer, while the remaining documents within the same evolving theme are treated as \texttt{non\_gold\_docs}. This explicit linkage enables rigorous supervision: a system must retrieve semantically relevant passages and identify the correct version aligned with the query's temporal attributes. Detailed prompts and instructions used for corpus generation are included in Appendix~\ref{sec:prompts}.
\begin{table}[t]
\centering
\caption{Benchmark comparison highlighting temporal complexity and domain specificity.}
\label{tab:bench_comparison}
\resizebox{\linewidth}{!}{%
    \begin{tabular}{@{}cc|cc|c@{}}
    \toprule
    \multicolumn{2}{c|}{\multirow{3}{*}{\textbf{\begin{tabular}[c]{@{}c@{}}Benchmark\\ Datasets\end{tabular}}}} &
      \multicolumn{2}{c|}{\textbf{Time Sensitivity}} &
      \multirow{3}{*}{\textbf{\begin{tabular}[c]{@{}c@{}}Domain \\  Specificity\end{tabular}}} \\ \cmidrule(lr){3-4}
    \multicolumn{2}{c|}{} &
      \multicolumn{1}{c|}{\textbf{\begin{tabular}[c]{@{}c@{}}\# Required\\  Time Types\end{tabular}}} &
      \textbf{\begin{tabular}[c]{@{}c@{}}Avg(\#Available \\ /\#Required)\end{tabular}} &
       \\ \midrule
    \multicolumn{1}{c|}{\multirow{4}{*}{\textbf{\benchname{}}}} &
      Company &
      \multicolumn{1}{c|}{2.16} &
      \heatG{100} & \heatR{4.20} \\
    \multicolumn{1}{c|}{} & Law        & \multicolumn{1}{c|}{2.04} & \heatG{100} & \heatR{4.83} \\
    \multicolumn{1}{c|}{} & Terms      & \multicolumn{1}{c|}{2.06} & \heatG{100} & \heatR{4.29} \\
    \multicolumn{1}{c|}{} & University & \multicolumn{1}{c|}{2.32} & \heatG{100} & \heatR{4.28} \\ \midrule
    \multicolumn{2}{c|}{TimeR$^{4}$}        & \multicolumn{1}{c|}{2.36} & \heatG{29.51} & \heatR{2.44} \\ \midrule
    \multicolumn{2}{c|}{CRAG}          & \multicolumn{1}{c|}{2.91} & \heatG{7.58}  & \heatR{3.04} \\ \midrule
    \multicolumn{2}{c|}{TS-Retriever}  & \multicolumn{1}{c|}{2.26} & \heatG{3.98}  & \heatR{2.81} \\ \midrule
    \multicolumn{2}{c|}{RealTime QA}   & \multicolumn{1}{c|}{2.01} & \heatG{95.53} & \heatR{3.12} \\ \midrule
    \multicolumn{2}{c|}{TimeQA}        & \multicolumn{1}{c|}{2.17} & \heatG{21.33} & \heatR{2.53} \\ \bottomrule
    \end{tabular}
} % end resizebox
\vspace*{-0.1cm}
\end{table}

\subsection{Multi-Dimensional Validation}
The third phase validates generated instances using GPT-4o, Claude-3.7-Sonnet, and Qwen2.5-72B as independent validators. We use multi-model agreement as a scalable consistency filter. Each validator assesses four criteria:
\begin{itemize}[leftmargin=9pt, noitemsep]
\item \textbf{Relevance:} Whether the query matches the intended category in Table~\ref{tab:query_category}.
\item \textbf{Derivability:} Whether the provided answer is fully supported by the annotated \texttt{gold\_docs}.
\item \textbf{Time Sensitivity:} Whether answering requires temporal reasoning, including the number of time types\,($Q_{IT}, Q_{ET}, D_{IT}, D_{ET}$) and the frequency of changes\,(yearly, monthly, daily, timely).
\item \textbf{Domain Specificity:} Whether the query and documents reflect professional style and terminology appropriate to the chosen domain.
\end{itemize}

\begin{table}[t]
    \centering
    \caption{Semantic overlap comparison across benchmark datasets. Higher BGE similarity indicates stronger semantic overlap among candidate documents.}
    \resizebox{\linewidth}{!}{%
        \begin{tabular}{@{}lcc@{}}
        \toprule
        \multicolumn{1}{c}{\textbf{Dataset}} &
        \multicolumn{1}{c}{\textbf{\#Queries}} &
        \multicolumn{1}{c}{\textbf{BGE Similarity}} \\
        \midrule
        \benchname{}-Company & 3000 & 0.71 \\
        \benchname{}-Law & 3000 & 0.61 \\
        \benchname{}-Terms of Service & 3000 & 0.73 \\
        \benchname{}-University & 3000 & 0.68 \\
        \midrule
        StreamingQA\,\cite{liska2022streamingqa} & 36376 & 0.35 \\
        CRAG\,\cite{yang2024crag} & 2705 & 0.36 \\
        RealTimeQA\,\cite{kasai2023realtime} & 1429 & 0.35 \\
        TS-Retriever\,\cite{wu2024time} & 3244 & 0.58 \\
        TimeR$^4$\,\cite{qian2024timer4} & 3237 & 0.34 \\
        \bottomrule
        \end{tabular}
    }
    \label{tab:semantic_overlap}
    \vspace{-0.2cm}
\end{table}

% the dataset quality by using three advanced LLMs---GPT-4o, Claude-3.7-Sonnet, and Qwen2.5-72B---each prompted independently to evaluate query-answer pairs, their alignment with supporting documents, and temporal validity. By combining judgments from multiple strong LLMs with complementary reasoning capabilities, we obtain validation reliability comparable to human assessment.
% Each query-answer pair is assessed across four criteria:
% \begin{itemize}[leftmargin=9pt, noitemsep]
% \item \textbf{Relevance:} Whether the query matches the intended category in Table~\ref{tab:query_category}.
% \item \textbf{Derivability:} Whether the provided answer is fully supported by the annotated \texttt{gold\_docs}.
% \item \textbf{Time Sensitivity:} Whether answering requires temporal reasoning, including the number of time types\,($Q_{IT}, Q_{ET}, D_{IT}, D_{ET}$) and the frequency of changes\,(yearly, monthly, daily, timely).
% \item \textbf{Domain Specificity:} Whether the query and documents reflect professional style and terminology appropriate to the chosen domain.
% \end{itemize}
%
% This validation confirms that the benchmark offers realistic, temporally grounded challenges. The annotation of multiple time types ensures that systems must incorporate diverse temporal signals, and the domain grounding ensures that questions reflect practical, regulation-driven scenarios where misinterpreting temporal validity may cause concrete regulatory or legal consequences. 
An instance is retained only when validators agree that the answer is supported by the annotated gold documents and requires temporal reasoning. This filtering reduces inconsistent instances while preserving examples that require clause-level temporal alignment. See Appendix~\ref{sec:validation_results} for validation results and Appendix~\ref{sec:prompts} for validation prompts.

\begin{table*}[t]
    \centering
    \caption{Retrieval performance on \benchname{}. Each cell reports baseline / \textbf{+\algname{}}.}
    \scriptsize
    \setlength{\tabcolsep}{2.5pt}
    \renewcommand{\arraystretch}{0.9}
    \resizebox{\linewidth}{!}{
    \begin{tabular}{@{}c|cccc|cccc@{}}
    \toprule
    \multirow{3}{*}{\textbf{Retriever}} &
    \multicolumn{4}{c|}{\textbf{Law}} &
    \multicolumn{4}{c}{\textbf{University}} \\
    \cmidrule(lr){2-5}\cmidrule(l){6-9}
    & \textbf{nDCG@10} & \textbf{MAP@10} & \textbf{Recall@10} & \textbf{Hit@10}
    & \textbf{nDCG@10} & \textbf{MAP@10} & \textbf{Recall@10} & \textbf{Hit@10} \\
    \midrule
    BM25 
    & .174/\textbf{.189} & .136/\textbf{.152} & .234/\textbf{.243} & .290/\textbf{.300}
    & .197/\textbf{.246} & .157/\textbf{.213} & .236/\textbf{.250} & .316/\textbf{.340} \\
    BGE-M3 
    & .172/\textbf{.179} & .136/\textbf{.145} & .229/\textbf{.231} & .280/\textbf{.283}
    & .173/\textbf{.213} & .142/\textbf{.187} & .196/\textbf{.213} & .265/\textbf{.290} \\
    NV-Embed-V2
    & .179/\textbf{.221} & .141/\textbf{.185} & .244/\textbf{.269} & .295/\textbf{.327}
    & .183/\textbf{.222} & .152/\textbf{.196} & .208/\textbf{.223} & .272/\textbf{.300} \\
    BGE-Gemma2
    & .183/\textbf{.224} & .144/\textbf{.188} & .249/\textbf{.274} & .302/\textbf{.334}
    & .181/\textbf{.216} & .151/\textbf{.191} & .204/\textbf{.217} & .267/\textbf{.288} \\
    \midrule
    \multirow{3}{*}{\textbf{Retriever}} &
    \multicolumn{4}{c|}{\textbf{Company}} &
    \multicolumn{4}{c}{\textbf{Terms of Service}} \\
    \cmidrule(lr){2-5}\cmidrule(l){6-9}
    & \textbf{nDCG@10} & \textbf{MAP@10} & \textbf{Recall@10} & \textbf{Hit@10}
    & \textbf{nDCG@10} & \textbf{MAP@10} & \textbf{Recall@10} & \textbf{Hit@10} \\
    \midrule
    BM25 
    & .226/\textbf{.227} & .182/\textbf{.183} & .218/\textbf{.219} & .344/\textbf{.345}
    & .113/\textbf{.134} & .094/\textbf{.115} & .111/\textbf{.123} & .170/\textbf{.190} \\
    BGE-M3 
    & .197/\textbf{.200} & .163/\textbf{.166} & .176/\textbf{.177} & .292/\textbf{.294}
    & .059/\textbf{.071} & .049/\textbf{.060} & .056/\textbf{.064} & .090/\textbf{.105} \\
    NV-Embed-V2
    & .195/\textbf{.228} & .161/\textbf{.197} & .175/\textbf{.179} & .289/\textbf{.312}
    & .062/\textbf{.078} & .050/\textbf{.067} & .065/\textbf{.073} & .099/\textbf{.115} \\
    BGE-Gemma2
    & .197/\textbf{.230} & .161/\textbf{.198} & \textbf{.183}/\textbf{.183} & .297/\textbf{.315}
    & .068/\textbf{.088} & .055/\textbf{.074} & .072/\textbf{.084} & .110/\textbf{.131} \\
    \bottomrule
    \end{tabular}
    }
    \label{tab:retrieval_results}
    \vspace{-0.2cm}
\end{table*}

\subsection{\benchname{} Analysis}
\label{sec:benchmark_analysis}

Table~\ref{tab:bench_comparison} compares \benchname{} with existing benchmarks in temporal coverage and domain specificity using Qwen2.5-72B. While prior datasets such as TimeQA\,\cite{chen2021dataset} and CRAG\,\cite{yang2024crag} only partially annotate temporal signals, \benchname{} provides comprehensive annotations across time types for clause-level temporal alignment. Its domain specificity makes \benchname{} practical for evaluating systems that retrieve the clause valid at query time.

To verify that \benchname{} captures overlapping evolution, we compute BGE similarity\,\cite{chen2024m3} among candidate documents within each evolving policy family. As shown in Table~\ref{tab:semantic_overlap}, \benchname{} exhibits substantially higher semantic overlap than existing time-sensitive QA benchmarks. This confirms that the benchmark cannot be solved by semantic matching alone and requires identifying the temporally valid version.

% contain few annotated time types or provide them only partially, which restricts the evaluation of queries requiring fine temporal reasoning. \benchname{} addresses this limitation by requiring multiple time types per query and guaranteeing complete annotation, ensuring supervision that enforces version-sensitive alignment.
% Furthermore, \benchname{} covers regulation-focused domains that undergo frequent amendments, where even minor changes in effective dates or provisions alter the correct answer. By reflecting overlapping-evolving environments and linking queries explicitly to supporting document versions, \benchname{} provides a realistic and demanding benchmark for testing retrieval systems designed to return the right version at the right time.

\section{Experimental Results}
\label{sec:evaluation}

\subsection{Experimental Setup}

\paragraph{Backbone Retrievers.}
We evaluate \algname{} on top of both sparse and dense retrievers, including BM25\,\cite{robertson1994some}, BGE-M3\,\cite{chen2024m3}, NV-Embed-V2\,\cite{lee2025nv}, and BGE-Gemma2\,\cite{chen2024m3}. \algname{} is applied as a temporal reranker on top of each fixed first-stage retriever.

\paragraph{Datasets.}
We use three benchmarks: \benchname{} for overlapping-evolving environments, TS-Retriever~\cite{wu2024time} for disjoint-evolving environments, and FiQA (BEIR)~\cite{thakur2021beir} for non-evolving environments. Details of these datasets are provided in Appendix~\ref{sec:dataset_details}.

% \begin{itemize}[leftmargin=9pt, noitemsep]
%     \item \textbf{\benchname{}}: Our newly constructed benchmark, comprising four regulation-focused domains: \textit{legal regulations}, \textit{university regulations}, \textit{company policy}, and \textit{terms of service}.
%     % \item \textbf{CRAG}~\cite{yang2024crag}: A general-purpose open-domain QA dataset widely used for evaluating RAG-based systems.
%     \item \textbf{TS-Retriever}~\cite{wu2024time}: A benchmark designed specifically for time-sensitive retrieval scenarios, built upon Wikipedia revisions and timestamped questions.
%     \item \textbf{FiQA (BEIR)}~\cite{thakur2021beir}: A financial-domain benchmark from the BEIR suite, containing user questions on stock market and financial regulations, used to test retrieval effectiveness.
%     % Each domain in \benchname{} includes queries and documents that demand precise temporal alignment, making them ideal for evaluating the temporal reasoning capability of retrieval systems.
% \end{itemize}

\paragraph{Retrieval Evaluation.}
We use four standard metrics: nDCG$@k$ for rank-sensitive relevance, MAP$@k$ for ranking quality, Recall$@k$ for relevant-document coverage, and Hit$@k$ for top-$k$ success.

\paragraph{Generation Evaluation.}
We assess downstream QA quality by comparing answers from baseline RAG and \algname{}-augmented RAG. For each query, both systems use the same generator, Llama-3.3-70B. Pairwise evaluation is conducted using GPT-5 as the judge model, which compares the two answers for correctness and temporal validity and reports win, tie, and lose rates for \algname{}.

\subsection{Retrieval Performance Comparison}
Results are presented for three types of benchmarks. 
$\alpha(Q)$ (\S\ref{sec:alpha_Q}) is set by the first heuristic for \S\ref{sec:overlapping_evolving_benchmark} and by the second heuristic for \S\ref{sec:disjoint_evolving_benchmark} and \S\ref{sec:general_RAG_benchmark}.

\subsubsection{Overlapping-Evolving Benchmark}
\label{sec:overlapping_evolving_benchmark}
We evaluate \algname{} on four \benchname{} domains: \textit{Law}, \textit{University}, \textit{Company}, and \textit{Terms of Service}. Table~\ref{tab:retrieval_results} shows that \algname{} improves retrieval across sparse and dense backbones. The largest gains appear in \textit{University} and \textit{Terms of Service}, with up to 24.8\% nDCG$@10$ improvement over BM25 on \textit{University} and 28.6\% over BGE-Gemma2 on \textit{Terms of Service}. Recall$@10$ and Hit$@10$ further confirm that temporal reranking improves both coverage and ranking accuracy. See Appendix~\ref{sec:additional_results} for extended results.

We additionally compare \algname{} with four stronger time-aware strategies. \emph{Temporal Query Rewriting} augments the original query with its temporal context at multiple granularities\,(timestamp, date, month, and year) before retrieval. \emph{Rule-based Effective-Date Filtering} applies hard temporal constraints to remove candidates whose effective periods are incompatible with the query. \emph{LLM-based Temporal Reranking} uses GPT-4o-mini to rerank retrieved candidates based on their textual content and temporal metadata. \emph{Oracle Temporal Filtering} uses gold temporal information to filter candidates and is included only as an upper bound.

\begin{table}[t]
    \centering
    \caption{Comparison with additional time-aware baselines on \benchname{}.}
    \resizebox{\linewidth}{!}{%
        \begin{tabular}{@{}l|l|cc@{}}
        \toprule
        \multicolumn{1}{c|}{\textbf{Method}} &
        \multicolumn{1}{c|}{\textbf{Backbone}} &
        \multicolumn{1}{c}{\textbf{nDCG@5}} &
        \multicolumn{1}{c}{\textbf{MAP@5}} \\
        \midrule
        Vanilla Retrieval
            & BM25   & 0.2087 & 0.1846 \\
        Vanilla Retrieval
            & BGE-M3 & 0.1784 & 0.1594 \\
        \midrule
        Temporal Query Rewriting
            & BM25   & 0.1486 & 0.1314 \\
        Temporal Query Rewriting
            & BGE-M3 & 0.1351 & 0.1205 \\
        Rule-based Effective-Date Filtering
            & BM25   & 0.1683 & 0.1500 \\
        Rule-based Effective-Date Filtering
            & BGE-M3 & 0.1386 & 0.1237 \\
        LLM-based Temporal Reranking
            & --     & 0.1948 & 0.1775 \\
        \midrule
        \rowcolor{gray!15}
        \algname{}
            & BM25   & \textbf{0.2821} & \textbf{0.2607} \\
        \rowcolor{gray!15}
        \algname{}
            & BGE-M3 & \textbf{0.2326} & \textbf{0.2154} \\
        \midrule
        Oracle Temporal Filtering
            & BM25   & 0.3323 & 0.3066 \\
        Oracle Temporal Filtering
            & BGE-M3 & 0.2771 & 0.2537 \\
        \bottomrule
        \end{tabular}
    }
    \label{tab:temporal_baselines}
\end{table}

As shown in Table~\ref{tab:temporal_baselines}, \algname{} consistently outperforms all deployable time-aware baselines under both sparse and dense retrieval. Temporal query rewriting merely injects temporal expressions into the query, while rule-based filtering applies hard date constraints. In contrast, \algname{} explicitly models temporal compatibility among query and document timestamps while preserving semantic relevance. The oracle results indicate that further gains remain possible with more accurate temporal information.

\subsubsection{Disjoint-Evolving Benchmark}
\begin{table}[t]
    \centering
    \caption{Retrieval performance comparison on the TS-Retriever dataset using nDCG, Recall, and Hit Rate.}
    \resizebox{\linewidth}{!}{
    \begin{tabular}{@{}c|ccc|ccc@{}}
    \toprule
    \multirow{4}{*}{\textbf{Methods}} &
      \multicolumn{6}{c}{\textbf{TS-Retriever Dataset}} \\
    \cmidrule(l){2-7}
     &
      \multicolumn{3}{c|}{\textbf{@3}} &
      \multicolumn{3}{c}{\textbf{@5}} \\
    \cmidrule(lr){2-4}\cmidrule(l){5-7}
     &
      \textbf{nDCG} &
      \textbf{Recall} &
      \multicolumn{1}{c|}{\textbf{Hit}} &
      \textbf{nDCG} &
      \textbf{Recall} &
      \textbf{Hit} \\
    \midrule
    TS-Retriever &
      0.539 & 0.406 & \multicolumn{1}{c|}{0.796} &
      0.547 & \textbf{0.527} & 0.865 \\
    \midrule
    BM25 &
      0.022 & 0.011 & \multicolumn{1}{c|}{0.032} &
      0.025 & 0.015 & 0.040 \\
    \rowcolor{gray!15}
    +\textbf{\algname{}} &
      0.641 & 0.334 & \multicolumn{1}{c|}{0.727} &
      0.679 & 0.391 & 0.817 \\
    \midrule
    SentenceBERT &
      0.392 & 0.190 & \multicolumn{1}{c|}{0.465} &
      0.425 & 0.254 & 0.557 \\
    \rowcolor{gray!15}
    +\textbf{\algname{}} &
      0.811 & 0.414 & \multicolumn{1}{c|}{0.857} &
      0.818 & 0.443 & 0.882 \\
    \midrule
    BGE-M3 &
      0.716 & 0.422 & \multicolumn{1}{c|}{0.796} &
      0.729 & \textbf{0.528} & 0.853 \\
    \rowcolor{gray!15}
    +\textbf{\algname{}} &
      \textbf{0.874} & \textbf{0.435} & \multicolumn{1}{c|}{\textbf{0.879}} &
      \textbf{0.875} & 0.451 & \textbf{0.891} \\
    \bottomrule
    \end{tabular}
    }
    \label{tab:tsretriever_results}
    \vspace{-0.3cm}
\end{table}

\label{sec:disjoint_evolving_benchmark}
Table~\ref{tab:tsretriever_results} reports results on TS-Retriever, which evaluates \textit{time-sensitive} retrieval. Across metrics, \algname{} consistently improves the baselines, showing that temporal alignment complements both lexical and semantic retrieval. These results confirm that \algname{} also adapts to disjoint-evolving scenarios by favoring temporally valid evidence over merely recent or lexically similar passages.

\subsubsection{General RAG Benchmark}
\label{sec:general_RAG_benchmark}
\begin{table}[t]
    \centering
    \caption{Retrieval results on the BEIR-FiQA dataset.}
    
    \resizebox{\linewidth}{!}{
    \begin{tabular}{@{}c|cccc@{}}
    \toprule
    \multirow{2.5}{*}{\textbf{Methods}} & \multicolumn{4}{c}{\textbf{BEIR-FiQA Dataset}}                            \\ \cmidrule(l){2-5} 
                                      & \textbf{nDCG@10} & \textbf{MAP@10} & \textbf{Recall@10} & \textbf{Hit@10} \\ \midrule
    BM25                              & 0.199            & 0.163           & 0.180              & 0.302           \\
    \textbf{+\algname{}}              & 0.199            & 0.163           & 0.180              & 0.302           \\ \midrule
    SentenceBERT                      & 0.485            & 0.419           & 0.439              & 0.653           \\
    \textbf{+\algname{}}              & 0.485            & 0.419           & 0.439              & 0.653           \\ \midrule
    BGE-M3                            & 0.729            & 0.466           & 0.660              & 0.907           \\
    \textbf{+\algname{}}              & 0.729            & 0.466           & 0.660              & 0.907           \\ \bottomrule
    \end{tabular}
    
    }
    \label{tab:fiqa_results}
    \vspace{-0.1cm}
\end{table}

To assess whether our temporal re-ranking mechanism interferes with retrieval in \emph{non}-temporal domains, we evaluate \algname{} on \textit{general RAG benchmarks} that are unrelated to time-dependent reasoning.
As shown in Table~\ref{tab:fiqa_results}, \algname{} preserves the original retriever performance because time-aware scoring becomes inactive without temporal context. This demonstrates that \algname{} preserves baseline effectiveness in static corpora while improving retrieval in time-critical settings.

\subsection{Generation Evaluation}
\begin{figure}[t]
    \centering
    \includegraphics[width=\linewidth]{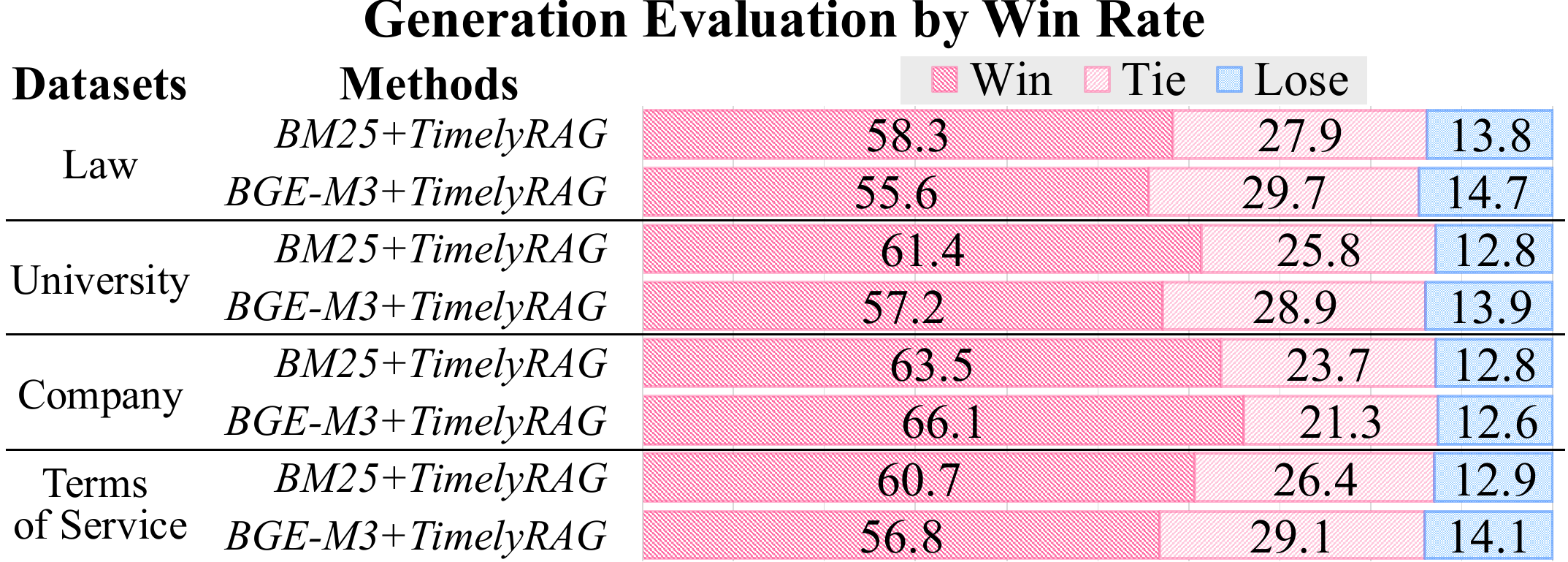}
    \caption{Generation performance of \algname{} compared with baseline RAG.}
    \label{fig:win_rate}
    \vspace{-0.2cm}
\end{figure}

Figure~\ref{fig:win_rate} shows that \algname{}-augmented generation consistently achieves higher win rates across all domains and retrievers. These results indicate that temporally valid retrieval improves evidence ranking and downstream answer reliability.

\subsection{Temporal Sensitivity Analysis}

\begin{figure}[t]
    \centering
    \includegraphics[width=\linewidth]{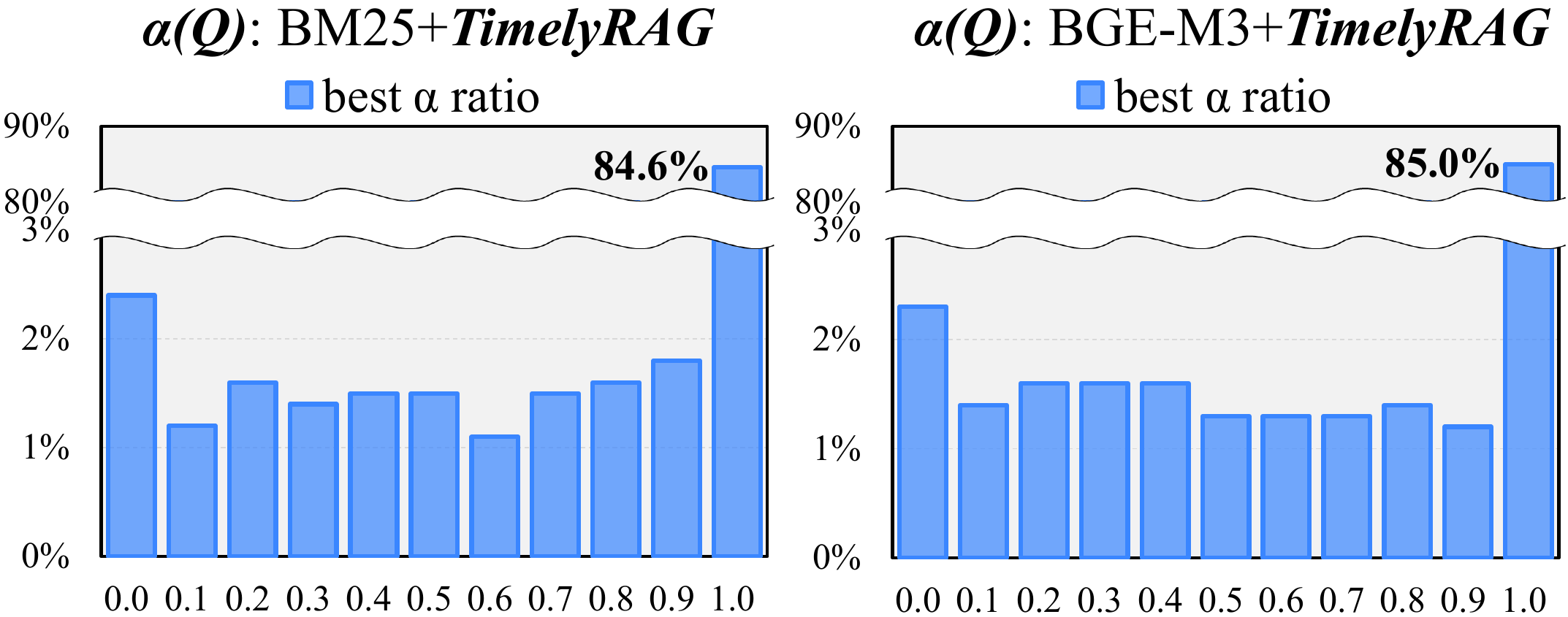}
    \caption{$\alpha(Q)$ distribution on the \textit{University} dataset. Higher $\alpha$ indicates stronger temporal importance.}
    \label{fig:alpha_distribution}
    \vspace{-0.0cm}
\end{figure}

Figure~\ref{fig:alpha_distribution} illustrates the distributions of optimal $\alpha(Q)$ values on the \benchname{}-\textit{University} dataset, where $\alpha(Q)$ controls the balance between semantic similarity and temporal alignment. In both sparse and dense retrievers, the optimal $\alpha(Q)$ values are heavily concentrated near 1.0 (84.6 \% and 85.0 \%), indicating that \benchname{} requires clause-level temporal validity rather than semantic similarity alone.
% revealing that most queries rely primarily on temporal consistency rather than semantic proximity. This strong skew toward high $\alpha(Q)$ confirms that the \benchname{} is highly time-critical, as correct retrieval requires clause-level validity matching—identifying the clause in effect at the query time.

\begin{table}[t]
    \centering
    \caption{Ablation results on the \benchname{}-\textit{University} dataset with the BM25 backbone.}
    
    \resizebox{\linewidth}{!}{
    \begin{tabular}{@{}c|cccc@{}}
    \toprule
    \multirow{2.5}{*}{\textbf{Variations}} & \multicolumn{4}{c}{\textbf{\benchname{}-University}}                      \\ \cmidrule(l){2-5} 
                                         & \textbf{nDCG@10} & \textbf{MAP@10} & \textbf{Recall@10} & \textbf{Hit@10} \\ \midrule
    Naive-Dist                           & 0.202            & 0.162           & 0.240              & 0.320           \\
    w/o $\mathbf{1}_{\text{has ET}(Q)}$                     & 0.205            & 0.166           & 0.238              & 0.319           \\
    w/o $\text{Gran}(Q)$                          & 0.206            & 0.168           & 0.237              & 0.318           \\ \midrule
    \textbf{\algname{}}                   & \textbf{0.246}   & \textbf{0.213}  & \textbf{0.250}     & \textbf{0.340}  \\ \bottomrule
    \end{tabular}
    }
    \label{tab:ablation}
    \vspace{-0.1cm}
\end{table}

\subsection{Ablation Study}
We analyze each component using simplified variants of our temporal formulation. The first applies all distance terms in Eq.~\eqref{eq:temporal_distance} with uniform weighting, ignoring causal constraints and \textit{fine-grained temporal relations}. The second removes the event-time indicator $\mathbf{1}_{\text{has ET}(Q)}$ in Eq.~\eqref{eq:final_alpha}, and the third omits the granularity term $\text{Gran}(Q)$. Table~\ref{tab:ablation} shows that query-document temporal alignment, event-time activation, and temporal granularity are effective for fine-grained time-critical retrieval. Full results are reported in Appendix~\ref{sec:expanded_ablation}.

\subsection{Computational Overhead}
\begin{table}[t]
    \centering
    \caption{Retrieval effectiveness and incremental temporal reranking overhead averaged across the four \benchname{} domains.}
    \resizebox{\linewidth}{!}{%
        \begin{tabular}{@{}l|c>{\columncolor{gray!15}}c|c>{\columncolor{gray!15}}c|c@{}}
        \toprule
        \multicolumn{1}{c|}{\textbf{Backbone}} &
        \multicolumn{2}{c|}{\textbf{nDCG@5}} &
        \multicolumn{2}{c|}{\textbf{MAP@5}} &
        \multicolumn{1}{c}{\textbf{Overhead}} \\
        & \textbf{Vanilla}
        & \textbf{+\algname{}}
        & \textbf{Vanilla}
        & \textbf{+\algname{}}
        & \textbf{ms/query} \\
        \midrule
        BM25
            & 0.2087 & \textbf{0.2821}
            & 0.1846 & \textbf{0.2607}
            & 2.15 \\
        BGE-M3
            & 0.1783 & \textbf{0.2326}
            & 0.1593 & \textbf{0.2156}
            & 2.64 \\
        \bottomrule
        \end{tabular}
    }
    \label{tab:runtime}
\end{table}
We measure the incremental query-time cost of \algname{} separately from the underlying retriever. Averaged across the four \benchname{} domains, in Table \ref{tab:runtime}, \algname{} adds only 2.15\,ms/query to BM25 and 2.64\,ms/query to BGE-M3 while improving nDCG@5. The reranking stage consists mainly of candidate sorting, temporal feature computation, and semantic--temporal mixing. For BGE-M3, document encoding is performed offline and excluded from query-time latency. The temporal module uses approximately $5.9\times10^{3}$ CPU-side operations per query, compared with $1.23\times10^{8}$ FLOPs/query for dense similarity computation. Overall, \algname{} provides substantial retrieval gains with \emph{little} incremental overhead.
% For reference, dense similarity computation with BGE-M3 requires approximately $1.23\times10^{8}$ FLOPs/query, whereas the temporal module performs approximately $5.9\times10^{3}$ CPU-side operations per query. Because the latter primarily consists of datetime comparisons and sorting, we report the operation count rather than treating it as directly comparable FLOPs.

\subsection{Candidate-Pool Sensitivity}
\begin{table}[t]
    \centering
    \caption{Sensitivity analysis of the first-stage candidate pool size $N$. Results are averaged across the four \benchname{} domains.}
    \resizebox{0.7\linewidth}{!}{%
        \begin{tabular}{@{}c|c|cc@{}}
        \toprule
        \multicolumn{1}{c|}{\textbf{Retriever}} &
        \multicolumn{1}{c|}{\boldmath$\mathbf{N}$} &
        \multicolumn{1}{c}{\textbf{nDCG@5}} &
        \multicolumn{1}{c}{\textbf{MAP@5}} \\
        \midrule
        \multirow{4}{*}{BM25}
            & 10  & 0.2646 & 0.2466 \\
            & 50  & 0.2821 & 0.2607 \\
            & 100 & \textbf{0.2830} & \textbf{0.2615} \\
            & 250 & 0.2820 & 0.2604 \\
        \midrule
        \multirow{4}{*}{BGE-M3}
            & 10  & 0.2176 & 0.2033 \\
            & 50  & 0.2328 & 0.2156 \\
            & 100 & 0.2342 & 0.2165 \\
            & 250 & \textbf{0.2353} & \textbf{0.2172} \\
        \bottomrule
        \end{tabular}
    }
    \label{tab:pool_size}
\end{table}
We vary the first-stage candidate pool size $N \in \{10,50,100,250\}$. As shown in Table~\ref{tab:pool_size}, increasing the pool from 10 to 50 candidates substantially improves retrieval quality for both backbones, whereas further increases provide only marginal gains. This saturation indicates that \algname{} benefits from sufficient candidate diversity but is relatively insensitive once the initial candidate pool becomes moderately large.
\section{Conclusion}
\label{sec:conclusion}

We introduce \algname{}, a retriever-agnostic framework for time-critical QA over overlapping-evolving documents. By combining semantic relevance with clause-level temporal compatibility, \algname{} distinguishes semantically similar versions with different temporal validity. We also construct \benchname{}, a controlled benchmark for regulation-heavy domains with successive amendments. Experiments show consistent retrieval gains across retrievers and improve downstream generation quality. Overall, reliable QA over evolving documents requires version-valid, clause-level temporal retrieval beyond document-level relevance.

% \section*{Limitations}

% This document does not cover the content requirements for ACL or any
% other specific venue.  Check the author instructions for
% information on
% maximum page lengths, the required ``Limitations'' section,
% and so on.

% \section*{Acknowledgments}

% This document has been adapted
% by Steven Bethard, Ryan Cotterell and Rui Yan
% from the instructions for earlier ACL and NAACL proceedings.

\section*{Acknowledgments}

This research was supported by the MSIT\,(Ministry of Science, ICT), Korea, under the Top-Tier AI Global HRD invitation program (RS-2025-25461932, 50\%) supervised by the IITP\,(Institute for Information \& Communications Technology Planning \& Evaluation), by the Korea Institute of Science and Technology Information\,(KISTI) in 2026 (No.\ (KISTI)K26L3M1C1, 40\%), aimed at developing KONI\,(KISTI Open Neural Intelligence), a large language model specialized in science and technology, and by Artificial intelligence industrial convergence cluster development project funded by the Ministry of Science and ICT\,(MSIT, Korea) \& Gwangju Metropolitan City (AICA-26-AE000008, 10\%).

\section*{Limitations}
While this work aims to tackle realistic retrieval scenarios under \textit{time-critical} and \textit{overlapping-evolving environments}, several limitations remain. First, \algname{} uses an interpretable temporal distance function; extending it to model complex temporal logic, causal relations, and multi-event dependencies remains future work. Second, as a reranking framework, \algname{} benefits from a first-stage retriever that can include relevant document versions in the candidate pool. Third, temporal signals can be harder to obtain in documents that do not follow explicit amendment or effective-date conventions. In contrast, regulation-heavy corpora often provide effective periods and amendment dates through metadata or formulaic text. \algname{} mitigates missing temporal signals through fallback and query-adaptive weighting. Finally, \benchname{} is synthetically constructed to control semantic overlap and temporal validity. Extending it with real-world versioned corpora is an important future direction.

\section*{Ethics Statement}
This work constructs a synthetic benchmark for time-critical QA in regulation-heavy domains. The dataset does not contain personal information, human-subject data, or offensive content. Since the benchmark is synthetically generated and validated with LLMs, it may still reflect biases or artifacts from the underlying models and prompts. We mitigate this risk through multi-model validation and release prompts and evaluation resources to support reproducibility. For human validation, two independent annotators from our research group evaluated 50 randomly sampled instances using the same validation criteria as the LLM-based evaluation. \algname{} is intended to improve the retrieval of temporally valid evidence, but it should not be used as a substitute for professional legal, policy, or regulatory advice.
We release \benchname{}, prompts, and evaluation scripts for research use under a permissive open-source license. Users should follow the license terms of any external models or datasets used for comparison.

\bibliography{7-Reference}

\clearpage
\appendix
\section{Temporal Granularity}
\label{sec:method_detail}

Temporal granularity determines the resolution at which time expressions are interpreted in both queries and documents.  
The granularity plays a key role in balancing semantic relevance and temporal compatibility through the gating function in Eq.~\eqref{eq:final_alpha}.  
A finer granularity (e.g., \textit{hour}-level) indicates a stronger temporal constraint, 
while coarser levels (e.g., \textit{year}-level) correspond to broader or more general temporal references.

\subsection{Final Granularity}
We define the granularity $\text{Gran}(Q)$ as an element of the ordered set 
$\mathcal{G}=\{\text{hour},\ \text{day},\ \text{month},\ \text{year}\}$, 
with an ordinal mapping $\phi:\mathcal{G}\to\{0,1,2,3\}$ defined as  
$\phi(\text{year})=0$, $\phi(\text{month})=1$, $\phi(\text{day})=2$, and $\phi(\text{hour})=3$.  
A larger ordinal thus corresponds to a finer temporal resolution.

To unify temporal resolution between queries and documents, 
we first take the finer of the two:
\[
\text{Gran}(Q)
= \max\nolimits_{\prec}\{\text{Gran}_{\text{query}},\,\text{Gran}_{\text{doc}}\},
\]
ensuring that temporal scoring always adopts the highest available precision.  
When the two differ significantly, we compute $\Delta=|g_q-g_d|$ for their integer levels 
$g_q,g_d\!\in\!\{0,1,2,3\}$ and pull one level toward the finer side if $\Delta\!\ge\!2$, 
avoiding abrupt transitions such as from \textit{year} directly to \textit{day}.

The resulting $\text{Gran}(Q)$ defines the temporal unit 
(\textit{hour}, \textit{day}, \textit{month}, or \textit{year}) 
used in the temporal distance computation of Eq.~\eqref{eq:temporal_distance}.  
As described in Eq.~\eqref{eq:final_alpha}, $\alpha(Q)$ balances semantic relevance and temporal compatibility 
based on two factors—the presence of explicit event times and the final temporal granularity.  
A finer granularity (larger $\text{Gran}(Q)$) leads to a higher $\alpha(Q)$, 
thus assigning greater weight to temporal compatibility relative to semantic similarity.

If $Q_{ET}$ is missing, $\text{Gran}(Q)$ defaults to \textit{month}, 
and insertion times substitute missing event terms.  
Missing document times fall back to $D_{IT}$ under the causality constraint $D_{IT}<Q_{IT}$.  
When multiple $Q_{ET}$ or $D_{ET}$ values exist, 
the pair minimizing Eq.~\eqref{eq:temporal_distance} is selected.

\begin{table}[t]
    \centering
    \caption{Multi-dimensional validation results on \benchname{} using the Qwen2.5-72B model.}
    \resizebox{\linewidth}{!}{
    \begin{tabular}{@{}ccccc@{}}
    \toprule
    Criteria                           & Law  & University & Company & Service Terms \\ \midrule
    \multicolumn{1}{c|}{\# Query}      & 3000 & 3000       & 3000    & 3000             \\
    \multicolumn{1}{c|}{Relevance (5)} & 4.98 & 4.90        & 4.88    & 4.94             \\
    \multicolumn{1}{c|}{\begin{tabular}[c]{@{}c@{}}Derivability\\  (\#Docs)\end{tabular}} & 3.45 & 3.05 & 2.82 & 3.09 \\ \bottomrule
    \end{tabular}
    }
    \label{tab:validation_results}
\end{table}

\subsection{Query-Side Estimation}
We detect the query-side granularity $\text{Gran}_{\text{query}}$ by parsing explicit temporal expressions 
from fine to coarse patterns:
\begin{align*}
\text{hour-level} &: \ \text{patterns of \texttt{YYYY-MM-DD HH:MM}},\\
\text{day-level}  &: \ \text{patterns of \texttt{YYYY-MM-DD}},\\
\text{month-level}&: \ \text{patterns of \texttt{YYYY-MM}},\\
\text{year-level} &: \ \text{patterns of \texttt{YYYY}}.
\end{align*}
If multiple patterns are matched, the finest level is selected.  
If no explicit time appears, we default to \textit{month}-level, which empirically balances coverage and stability.

\subsection{Document-Side Estimation}
Let $\mathcal{C}(Q)\subseteq\mathcal{D}$ be the semantic top-$N$ candidate documents for query $Q$.  
We estimate the predominant temporal resolution among these candidates by analyzing their insertion and event timestamps $(D_{IT}, D_{ET})$.  
Specifically, we compute three proportions:  
$p_{\text{hour}}$, $p_{\text{day}}$, and $p_{\text{month}}$, 
representing the fractions of documents that (i) contain nonzero hour or minute values, 
(ii) have day components other than 1, and (iii) have month components other than 1, respectively.

The document-side granularity $\text{Gran}_{\text{doc}}$ is then determined by the finest level 
whose proportion exceeds a predefined threshold $\tau$ (set to $\tau=0.7$ in our experiments):
\[
\text{Gran}_{\text{doc}}=
\begin{cases}
\text{hour},  & \text{if }p_{\text{hour}} \ge \tau,\\
\text{day},   & \text{else if }p_{\text{day}} \ge \tau,\\
\text{month}, & \text{else if }p_{\text{month}} \ge \tau,\\
\text{year},  & \text{otherwise}.
\end{cases}
\]
This identifies the dominant temporal resolution in the retrieved candidates, 
which is then combined with the query-side estimate to obtain the final $\text{Gran}(Q)$.

\begin{table*}[t]
    \centering
      \caption{Additional results of retrieval performance comparison on the \benchname{} dataset.}
    \resizebox{\linewidth}{!}{
    \begin{tabular}{@{}c|cccc|cccc@{}}
    \toprule
    \multirow{2}{*}{\textbf{Methods}} &
      \textbf{nDCG@10} &
      \textbf{MAP@10} &
      \textbf{Recall@10} &
      \textbf{Hit@10} &
      \textbf{nDCG@10} &
      \textbf{MAP@10} &
      \textbf{Recall@10} &
      \textbf{Hit@10} \\ \cmidrule(l){2-9} 
                     & \multicolumn{4}{c|}{\textbf{Law}}                                     & \multicolumn{4}{c}{\textbf{University}}                               \\ \midrule
    TS-Retriever     & 0.0188          & 0.0162          & 0.0278          & 0.0317          & 0.0301          & 0.0241          & 0.0409          & 0.0557          \\ \midrule
    DPR              & 0.0007          & 0.0005          & 0.0009          & 0.0013          & 0.0001          & 0.0001          & 0.0002          & 0.0003          \\
    \rowcolor{gray!15}
    +\textbf{\algname{}}      & 0.0009          & 0.0007          & 0.0013          & 0.0017          & 0.0002          & 0.0001          & 0.0003          & 0.0007          \\ \midrule
    Contriever       & 0.0023          & 0.0015          & 0.0040          & 0.0047          & 0.0044          & 0.0036          & 0.0042          & 0.0070          \\
    \rowcolor{gray!15}
    +\textbf{\algname{}}      & 0.0026          & 0.0019          & 0.0042          & 0.0050          & 0.0056          & 0.0047          & 0.0053          & 0.0090          \\ \midrule
    ColBERT          & 0.0350          & 0.0297          & 0.0376          & 0.0520          & 0.0812          & 0.0705          & 0.0748          & 0.1130          \\
    \rowcolor{gray!15}
    +\textbf{\algname{}}      & \textbf{0.0410} & \textbf{0.0360} & \textbf{0.0410} & \textbf{0.0560} & \textbf{0.0972} & \textbf{0.0882} & \textbf{0.0810} & \textbf{0.1240} \\ \midrule
    \textbf{Methods} & \multicolumn{4}{c|}{\textbf{Company}}                                 & \multicolumn{4}{c}{\textbf{Terms of Service}}                         \\ \midrule
    TS-Retriever     & 0.0205          & 0.0159          & 0.0241          & 0.0427          & 0.0375          & 0.0301          & 0.0517          & 0.0710          \\ \midrule
    DPR              & 0.0004          & 0.0004          & 0.0002          & 0.0007          & 0.0006          & 0.0004          & 0.0004          & 0.0010          \\
    \rowcolor{gray!15}
    +\textbf{\algname{}}      & 0.0006          & 0.0004          & 0.0003          & 0.0010          & 0.0012          & 0.0008          & 0.0014          & 0.0027          \\ \midrule
    Contriever       & 0.0034          & 0.0024          & 0.0030          & 0.0067          & 0.0031          & 0.0024          & 0.0029          & 0.0057          \\
    \rowcolor{gray!15}
    +\textbf{\algname{}}      & 0.0041          & 0.0030          & 0.0030          & 0.0077          & 0.0047          & 0.0036          & 0.0042          & 0.0083          \\ \midrule
    ColBERT          & 0.0622          & 0.0545          & 0.0437          & 0.0840          & 0.0555          & 0.0478          & 0.0480          & 0.0787          \\
    \rowcolor{gray!15}
    +\textbf{\algname{}}      & \textbf{0.0722} & \textbf{0.0656} & \textbf{0.0461} & \textbf{0.0910} & \textbf{0.0651} & \textbf{0.0576} & \textbf{0.0525} & \textbf{0.0880} \\ \bottomrule
    \end{tabular}
    }
    \label{tab:additional_results}
\end{table*}

\section{Multi-Dimensional Validation Results}
\label{sec:validation_results}

Using the Qwen2.5-72B model, we evaluated \benchname{} datasets along four key dimensions—time sensitivity, domain specificity, relevance, and derivability.
The results for time sensitivity and domain specificity are presented earlier in Table~\ref{tab:bench_comparison}, while Table~\ref{tab:validation_results} reports the remaining results for relevance and derivability.
Relevance, rated on a 5-point scale, measures how closely each query–answer–document set aligns with its assigned category in Table~\ref{tab:query_category} (1: irrelevant, 5: highly relevant).
Across all domains, \benchname{} achieves an average relevance score of 4.93, indicating that the dataset is well aligned with time-critical categories.
Derivability represents the average number of passages required to answer each query.
On average, 3.10 passages are needed to derive a correct answer, reflecting the temporal complexity of consulting multiple revised versions to identify clauses valid at the query time.

\section{Additional Experimental Results}
\label{sec:additional_results}

\paragraph{Hardware Configuration.} All experiments are conducted on a server equipped with an NVIDIA RTX 6000 Ada Generation.

\subsection{Backbone Retrievers}
We employ the following backbone retrievers in our evaluation:
\begin{itemize}[leftmargin=9pt, noitemsep]
    \item \textbf{BM25}~\cite{robertson1994some}: a classical lexical model based on term frequency and inverse document frequency
    \item \textbf{BGE-M3}~\cite{chen2024m3}: a multi-function embedding model supporting dense, sparse, and multi-vector retrieval
    \item \textbf{Sentence-BERT}~\cite{reimers-gurevych-2019-sentence}: semantically meaningful sentence embeddings via siamese BERT networks
    \item \textbf{DPR}~\cite{karpukhin2020dense}: a standard supervised dense retriever trained on question–passage
    \item \textbf{Contriever}~\cite{izacard2022unsupervised}: learning dense representations in a fully unsupervised manner
    \item \textbf{ColBERT}~\cite{santhanam-etal-2022-colbertv2}: performing late-interaction scoring between query and document tokens for fine-grained semantic matching
\end{itemize}
All retrievers are tested both in their vanilla forms and with our temporal re-ranking module \algname{}, which adjusts ranking scores according to semantic–temporal compatibility.

\subsection{Datasets}
\label{sec:dataset_details}

\paragraph{\benchname{}.}
Our newly constructed benchmark comprises four regulation-focused domains: \textit{legal regulations}, \textit{university regulations}, \textit{company policy}, and \textit{terms of service}.

\paragraph{TS-Retriever.}
TS-Retriever~\cite{wu2024time} is a benchmark designed for time-sensitive retrieval scenarios, built upon Wikipedia revisions and timestamped questions.

\paragraph{FiQA (BEIR).}
FiQA (BEIR)~\cite{thakur2021beir} is a financial-domain benchmark from the BEIR suite, containing user questions on stock market and financial regulations.

\subsection{Results and Analysis}
Table~\ref{tab:additional_results} shows retrieval performance on the four subsets of \benchname{}.  
Across all backbones, the absolute scores of baseline retrievers are strikingly low, revealing the inherent difficulty of \benchname{}.  
Unlike conventional RAG datasets, where lexical or semantic similarity alone suffices, documents in \benchname{} exhibit extremely high semantic overlap across temporal versions.  
Many passages differ only in clause-level details—such as validity periods, effective dates, or scope restrictions—making temporal reasoning indispensable for correct retrieval.

When equipped with \algname{}, every retriever demonstrates consistent improvement across all metrics.
Even simple dense retrievers like DPR and Contriever benefit from our temporal re-ranking, indicating that \algname{} effectively captures fine-grained clause-level temporal distinctions beyond semantic similarity. These results confirm that \benchname{} is a realistic and challenging benchmark that cannot be solved by semantic matching alone.  
Accurate retrieval requires identifying the clause valid at the query time, and \algname{} provides this essential temporal discrimination capability.

\subsection{Human Validation}
\begin{table}[t]
    \centering
    \caption{Pilot human evaluation on 50 randomly sampled \benchname{} queries compared with full-dataset LLM validation.}
    \resizebox{\linewidth}{!}{%
        \begin{tabular}{@{}l|cccc@{}}
        \toprule
        \multicolumn{1}{c|}{\textbf{Evaluator}} &
        \multicolumn{1}{c}{\textbf{Time Types}} &
        \multicolumn{1}{c}{\textbf{Domain Spec.}} &
        \multicolumn{1}{c}{\textbf{Relevance}} &
        \multicolumn{1}{c}{\textbf{Derivability}} \\
        \midrule
        Human Evaluation
            & 2.73 & 4.38 & 4.74 & 3.39 \\
        LLM Validation
            & 2.15 & 4.40 & 4.93 & 3.10 \\
        \bottomrule
        \end{tabular}
    }
    \label{tab:human_validation}
\end{table}

We additionally conduct a pilot human evaluation on 50 randomly sampled \benchname{} queries. Two independent annotators from our research
group evaluate the samples using the same validation dimensions as our automatic pipeline\,(required temporal types, domain specificity,
relevance, and derivability). As shown in Table~\ref{tab:human_validation}, human judgments exhibit a trend consistent with the full-dataset LLM validation, with both evaluations indicating high relevance and domain specificity. We observe no substantial discrepancy between human and automatic judgments on this pilot sample.

\subsection{Category-Wise Analysis}

\begin{table}[t]
    \centering
    \caption{Per-category retrieval performance on the Company dataset (nDCG@10).}
    \resizebox{\linewidth}{!}{%
        \begin{tabular}{@{}l|c>{\columncolor{gray!15}}c|c>{\columncolor{gray!15}}c@{}}
        \toprule
        \multicolumn{1}{c|}{\textbf{Category}} &
        \multicolumn{1}{c}{\textbf{BM25}} &
        \multicolumn{1}{c|}{\textbf{+\algname{}}} &
        \multicolumn{1}{c}{\textbf{BGE-M3}} &
        \multicolumn{1}{c}{\textbf{+\algname{}}} \\
        \midrule
        Multi-Version Policy Rules & 0.222 & \textbf{0.225} & 0.171 & \textbf{0.181} \\
        Legacy Exception Rules & 0.263 & \textbf{0.271} & 0.246 & \textbf{0.246} \\
        Policy Name/Code Revisions & 0.201 & \textbf{0.214} & 0.155 & \textbf{0.156} \\
        Emergency/Temporary Rules & 0.287 & \textbf{0.297} & 0.300 & \textbf{0.304} \\
        Historical, Retrospective Fact & \textbf{0.186} & 0.157 & \textbf{0.133} & 0.130 \\
        Implicit Time Context & 0.197 & \textbf{0.199} & 0.176 & \textbf{0.182} \\
        \bottomrule
        \end{tabular}
    }
    \label{tab:category_analysis}
\end{table}
We analyze retrieval performance across the six query categories in Table~\ref{tab:query_category}. This analysis shows where temporal reranking is most beneficial and whether \algname{} remains effective in implicit temporal contexts, where the relevant time must be inferred rather than matched directly.

Table~\ref{tab:category_analysis} reports per-category nDCG@10 on the Company dataset. \algname{} improves or preserves performance in most categories, including Implicit Time Context, where explicit temporal cues are weak. The only clear degradation appears in Historical, Retrospective Fact, suggesting that retrospective queries require more precise handling of past validity intervals.

\subsection{Expanded Ablation}
\label{sec:expanded_ablation}
\begin{table}[t]
    \centering
    \caption{Expanded ablation study across \benchname{} domains and retrievers\,(nDCG@5).}
    \resizebox{\linewidth}{!}{%
        \begin{tabular}{@{}cc|ccc>{\columncolor{gray!15}}c@{}}
        \toprule
        \multicolumn{1}{c}{\textbf{Domain}} &
        \multicolumn{1}{c|}{\textbf{Retriever}} &
        \multicolumn{1}{c}{\textbf{Naive-Dist}} &
        \multicolumn{1}{c}{\textbf{w/o ET}} &
        \multicolumn{1}{c}{\textbf{w/o Gran.}} &
        \multicolumn{1}{c}{\textbf{\algname{}}} \\
        \midrule
        \multirow{2}{*}{Company}
            & BM25   & 0.203 & 0.214 & 0.221 & \textbf{0.227} \\
            & BGE-M3 & 0.181 & 0.189 & 0.195 & \textbf{0.200} \\
        \midrule
        \multirow{2}{*}{Law}
            & BM25   & 0.162 & 0.174 & 0.183 & \textbf{0.189} \\
            & BGE-M3 & 0.154 & 0.164 & 0.173 & \textbf{0.179} \\
        \midrule
        \multirow{2}{*}{Terms of Service}
            & BM25   & 0.113 & 0.123 & 0.129 & \textbf{0.134} \\
            & BGE-M3 & 0.058 & 0.064 & 0.068 & \textbf{0.071} \\
        \midrule
        \multirow{2}{*}{University}
            & BM25   & 0.202 & 0.205 & 0.206 & \textbf{0.246} \\
            & BGE-M3 & 0.184 & 0.197 & 0.206 & \textbf{0.213} \\
        \bottomrule
        \end{tabular}
    }
    \label{tab:expanded_ablation}
\end{table}
Table~\ref{tab:expanded_ablation} extends the ablation study to all four domains and both sparse and dense retrievers. \algname{} consistently outperforms Naive-Dist and the variants without event-time activation\,(\textbf{w/o ET}) or temporal-granularity modeling \,(\textbf{w/o Gran.}) in all eight settings, demonstrating that the benefits of the proposed temporal modeling are not limited to a particular domain or retrieval backbone.

\section{Prompts}
\label{sec:prompts}

\begin{table*}[t]
\centering
\small
\caption{Time-critical Korean QA and corpus benchmark generation prompt.}
\label{tab:prompt_qa_corpus}
\resizebox{\textwidth}{!}{%
    \begin{tabular}{@{}p{0.05\textwidth}p{0.95\textwidth}@{}}
    \toprule
    \textbf{Step} & \textbf{Instruction} \\
    \midrule
    1. &
    \textbf{Role and Purpose}: You are an expert dataset builder for Korean corporate HR policies in a retrieval-augmented generation\,(RAG) system. Your task is to generate a realistic, time-sensitive benchmark sample that reflects how internal company regulations change across time and how those changes affect the answerability of a given query. The output must be in \textbf{Korean} and follow the structure below. \\
    \midrule
    2. &
    \textbf{Components to Include}: Each sample must include the following fields:
    \begin{itemize}
        \item \texttt\{category\}: Choose one from the following: \{\textit{category}\}
        \item \texttt\{query\_ko\}: A specific, realistic, and time-sensitive HR-related question that could be asked by an employee or HR manager.
        \begin{itemize}
            \item Avoid vague questions like ``what changed''.
            \item Vary queries by roles, years, departments, contract types, or policy situations.
            \item Incorporate time constraints, past usage, or eligibility.
            \item Ensure the question is answerable only with the correct versions of policy documents.
        \end{itemize}
        \item \texttt\{query\_insert\_time\}: A realistic past timestamp in the format \texttt\{YYYY-MM-DD HH:MM:SS\}.
        \item \texttt\{answer\}: A correct and complete answer in Korean, supported entirely by one or more of the documents.
        \item \texttt\{docs\_ko\}: A list of 10 internal policy passages.
        \begin{itemize}
            \item Each passage must be in Korean.
            \item Share a common policy topic.
            \item Content must vary to simulate time-based revisions.
            \item Each must include a \texttt\{text\} and \texttt\{doc\_insert\_time\}.
        \end{itemize}
        \item \texttt\{gold\_docs\}: List of 0-based indices in \texttt\{docs\_ko\} that support the answer.
    \end{itemize} \\
    \midrule
    3. &
    \textbf{Constraints}:
    \begin{itemize}
        \item \textbf{Do not} repeat previous queries.
        \item Avoid boilerplate phrasing.
        \item Each question must be \textbf{distinct}.
        \item Return \textbf{only} the JSON object in raw form.
        \item No explanations, no markdown, no formatting.
    \end{itemize} \\
    \midrule
    4. &
    \textbf{Output Format (JSON)}:
    \begin{quote}
    \ttfamily
        \begin{tabular}{@{}l@{}}
        \{ \\
        \ \ \ "category": "\{category\}", \\
        \ \ \ "query\_insert\_time": "YYYY-MM-DD HH:MM:SS", \\
        \ \ \ "query\_ko": "...", \\
        \ \ \ "answer": "...", \\
        \ \ \ "docs\_ko": [ ...,\\
        \ \ \ \ \ \{ "text": "...", "doc\_insert\_time": "YYYY-MM-DD HH:MM:SS" \}, \\
        \ \ \ \ \ \{ "text": "...", "doc\_insert\_time": "YYYY-MM-DD HH:MM:SS" \}, \\
        \ \ \ \ \ \{ "text": "...", "doc\_insert\_time": "YYYY-MM-DD HH:MM:SS" \}, \\
        \ \ \ \ \ \{ "text": "...", "doc\_insert\_time": "YYYY-MM-DD HH:MM:SS" \}, \\
        \ \ \ \ \ \{ "text": "...", "doc\_insert\_time": "YYYY-MM-DD HH:MM:SS" \}, \\
        \ \ \ \ \ \ ... \\
        \ \ \ ], \\
        \ \ \ "gold\_docs": "..." \\
        \} \\
        \end{tabular}
    \end{quote}
    \\
    \bottomrule
    \end{tabular}%
}
\end{table*}

\begin{table*}[t]
\centering
\small
\caption{Validation prompt used to annotate time-critical QA samples across four dimensions.}
\label{tab:prompt_validation}
\resizebox{\textwidth}{!}{%
    \begin{tabular}{@{}p{0.05\textwidth}p{0.95\textwidth}@{}}
    \toprule
    \textbf{Step} & \textbf{Instruction} \\
    \midrule
    1. & \textbf{Role and Goal}: You are a professional Korean benchmark evaluator for time-sensitive document QA (within a RAG framework). Your task is to evaluate the provided QA sample across four validation dimensions. \\
    \midrule
    2. & \textbf{Category}: Explicitly note the QA topic domain (e.g., HR, legal, university regulations). \\
    \midrule
    3. & \textbf{Time Type Definitions}:
    \begin{itemize}
        \item \texttt{QIT} – Query Insert Time: When the user created the query.
        \item \texttt{QET} – Query Event Time: Temporal expressions mentioned in the query.
        \item \texttt{DIT} – Document Insert Time: When the document was added to the database.
        \item \texttt{DET} – Document Event Time: Time expressions mentioned inside the document.
    \end{itemize} \\
    \midrule
    4. & \textbf{Evaluation Dimensions}:
    \begin{itemize}
        \item \textbf{Relevance (1–5)}: Does the query match the designated category?
        \item \textbf{Derivability}: Identify the indices of documents required to derive the answer. Return as \texttt{supporting\_docs}.
        \item \textbf{Time Sensitivity}:
        \begin{itemize}
            \item Identify which time types are involved (\texttt{QIT}, \texttt{QET}, \texttt{DIT}, \texttt{DET}).
        \end{itemize}
        \item \textbf{Domain Specificity (1–5)}: Rate how professional or domain-specific the language is.
    \end{itemize} \\
    \midrule
    5. & 
    \textbf{Output Format (JSON)}:
    \begin{quote}
    \ttfamily
        \begin{tabular}{@{}l@{}}
        \{ \\
        \ \ \ "relevance": \{ \\
        \ \ \ \ \ \ "score": 1-5, \\
        \ \ \ \ \ \ "reason": "..." \\
        \ \ \ \}, \\
        \ \ \ "derivability": \{ \\
        \ \ \ \ \ \ "supporting\_docs": [0, 1, ...], \\
        \ \ \ \ \ \ "reason": "..." \\
        \ \ \ \}, \\
        \ \ \ "time\_sensitivity": \{ \\
        \ \ \ \ \ \ "available\_time\_types": ["QIT", "QET", "DIT", "DET"], \\
        \ \ \ \ \ \ "required\_time\_types": ["QIT", "QET", "DIT", "DET"], \\
        \ \ \ \ \ \ "reason": "..." \\
        \ \ \ \}, \\
        \ \ \ "domain\_specificity": \{ \\
        \ \ \ \ \ \ "score": 1-5, \\
        \ \ \ \ \ \ "reason": "..." \\
        \ \ \ \} \\
        \}
        \end{tabular}
    \end{quote} 
    \\
    \bottomrule
    \end{tabular}%
}
\end{table*}

\subsection{Benchmark Dataset Generation Prompt}

Table~\ref{tab:prompt_qa_corpus} presents the prompt template used for generating time-sensitive Korean QA samples in regulation-intensive domains. The prompt is carefully designed to guide the construction of realistic query-answer pairs that reflect temporal variability in corporate policies. It specifies structured requirements for each component---such as query timestamp, semantically overlapping but temporally distinct documents, and gold document labels---ensuring consistency and temporal grounding. This template plays a critical role in the creation of high-quality benchmark data for evaluating time-aware retrieval and generation systems.

\subsection{Benchmark Validation Prompt}
Table~\ref{tab:prompt_validation} presents the prompt template used for validating each QA instance in \benchname{}. This validation process is designed to assess multiple dimensions critical for time-sensitive QA evaluation. Specifically, annotators are asked to score the query's relevance to its category, identify which documents are necessary to derive the correct answer, determine which temporal types\,(e.g., $Q_{IT}, Q_{ET}, D_{IT}, D_{ET}$) are involved, and describe how frequently the answer may change over time. Additionally, the prompt guides annotators to evaluate the level of domain-specific language used in both the query and the documents. By standardizing this evaluation framework, the prompt ensures consistent, high-quality validation across the dataset.

\end{document}